\documentclass[journal]{IEEEtran}
\usepackage{cite}
\usepackage{graphicx}
\usepackage{amssymb,amsmath,bm}
\usepackage[colorlinks,linkcolor=black,anchorcolor=black,citecolor=black,urlcolor=black]{hyperref}
\usepackage{hyperref}
\usepackage{multirow}
\usepackage{amsfonts}
\usepackage{url}
\usepackage{algorithm}
\usepackage{algpseudocode}
\usepackage{xcolor}
\usepackage{multirow}
\usepackage{color}
\usepackage{tabularx}
\usepackage{color}
\usepackage{algorithm}
\usepackage{booktabs}
\usepackage{multirow}
\usepackage{array}
\usepackage{algpseudocode}
\usepackage{diagbox}
\usepackage{bbding}
\usepackage{comment}
\usepackage{enumitem}
\usepackage{enumitem}
\colorlet{blue}{black}
\setlist[itemize]{leftmargin=*, itemindent=0pt}
\ifCLASSINFOpdf
\else
\fi
\begin{document}
%
\title{Ultra-Low-Bitrate Mel-Spectrogram-based Neural Speech Coding with Flow-Matching-based Refinement and Vocoding-driven Reconstruction}
%
%
%

\author{Hui-Peng~Du, Yang~Ai,~\IEEEmembership{Member,~IEEE}, Xiao-Hang~Jiang, Yuan Tian,~Zhen-Hua~Ling,~\IEEEmembership{Senior Member,~IEEE}
\thanks{This work was funded by the National Nature Science Foundation of China under Grant 62301521. (Corresponding author: Yang~Ai)}
\thanks{Hui-Peng Du, Yang Ai, Xiao-Hang Jiang, Yuan Tian and Zhen-Hua Ling are with the National Engineering Research Center of Speech and Language Information Processing, University of Science and Technology of China, Hefei, 230027, China (e-mail: redmist@mail.ustc.edu.cn, yangai@ustc.edu.cn, \{jiang\_xiaohang, ytian1507\}@mail.ustc.edu.cn, zhling@ustc.edu.cn).}
}
%
%

\markboth{}%
{Shell \MakeLowercase{\textit{et al.}}: Bare Demo of IEEEtran.cls for Journals}
%



\maketitle

\begin{abstract}
Ultra-low-bitrate speech coding is pivotal for bandwidth-constrained communication and deep compression, yet maintaining naturalness and speaker identity at such extreme bit budgets remains challenging due to pronounced information loss and quantization instability.
To this end, we propose FMelCodec, an ultra-low-bitrate neural speech codec in the mel-spectrogram domain, cast as a three-stage coding–refinement–reconstruction (CRR) framework that can operate at as low as 250 bps.
In the CRR framework, the front-end mel-spectrogram coding stage employs a highly aggressive 640$×\times$  compression/decompression encoder–decoder structure with a single 1024-entry VQ codebook, coupled with an online clustering strategy that reassigns underused codewords to prevent codebook collapse and preserve codebook diversity.
The subsequent conditional flow matching (CFM)-based mel-spectrogram refinement stage leverages a lightweight velocity-field estimator and CFM-based solver to refine the codec-degraded mel-spectrogram produced by the preceding decoder, and adopts a self-consistency training scheme that supports fewer iterative inference steps for the purpose of reducing computational overhead.
Finally, the vocoding-driven waveform reconstruction stage employs a HiFi-GAN vocoder to faithfully reconstruct waveform from the refined mel-spectrogram.
Experiments conducted on two datasets spanning two sampling rates show that, under ultra-low-bitrate constraints of 250 bps for 16 kHz and 750 bps for 48 kHz, both objective and subjective evaluations consistently demonstrate that FMelCodec achieves higher speech reconstruction quality and speaker similarity, while incurring lower computational and model complexity.

\end{abstract}


%
\IEEEpeerreviewmaketitle

\section{Introduction}

\IEEEPARstart{S}{peech} codecs transform speech into compact discrete representations and reconstruct intelligible, natural-sounding speech from them.
They have long been essential for speech communication and compression \cite{valin2013high,dietz2015overview,kleijn2021generative}, and in recent years have begun to emerge as a promising component in a range of downstream speech {applications} 
{\cite{11303581,li2025speech,10842513,chen2024vall,defossez2024moshi}}.
Bitrate is a central metric for assessing speech codecs and is often a decisive factor in real-world deployment. 
In bandwidth-constrained settings such as satellite links or communication between low-resource devices, the available bitrate can be severely limited, quickly becoming the bottleneck that determines whether a codec is viable at all. 
Yet pushing bitrate lower requires increasingly aggressive compression and heightens the risk of removing perceptually salient information, rendering high-quality reconstruction under ultra-low-bitrate budgets a long-standing and fundamental challenge for speech coding. 

Early standardized speech codecs such as Opus~\cite{valin2013high} and EVS~\cite{dietz2015overview} are built on hand-crafted digital signal processing (DSP) pipelines.
For 16-kHz speech coding, standardized DSP codecs typically operate at moderate bitrates on the order of 6$\sim$20 kbps, while pushing the rate further down generally leads to a pronounced drop in timbre fidelity and speaker-related cues.
This bitrate–quality trade-off has spurred the development of neural speech codecs, which leverage powerful neural modeling and data-driven learning to acquire more expressive representations and quantization strategies, thereby maintaining high speech quality at reduced bitrates.

Waveform-domain SoundStream~\cite{zeghidour2021soundstream} and Encodec~\cite{defossez2023high} are pioneering neural codecs that substantially improve reconstruction quality at reduced bitrate over traditional DSP codecs by combining causal convolutional encoder--decoder backbones with learned residual vector quantization (RVQ) and adversarial training~\cite{goodfellow2014generative}. 
As a result, RVQ has emerged as a widely adopted and well-regarded quantization scheme in modern neural codecs, in which multiple VQ codebooks are cascaded and each stage quantizes the residual from the preceding one.
Building on these pioneers, subsequent waveform-domain studies \cite{kumar2024high,xin2024bigcodec} have further advanced the field by introducing increasingly sophisticated architectures to improve speech reconstruction quality. 
However, these codecs still analyze and reconstruct speech directly in the waveform domain, which typically requires deep stacks of downsampling and upsampling layers and leads to higher model and computational complexity, making them less amenable to resource-constrained, non-parallelizable edge deployment.

To alleviate the complexity burden, recent research has increasingly turned toward spectral-domain neural speech codecs \cite{ai2024apcodec,jiang2024mdctcodec,jiang2025streamable,langman2024spectral}. 
This line of work redirects speech codecs from analyzing and reconstructing raw waveforms to operating on spectral representations, offering a distinct alternative to waveform-domain approaches. 
For example, APCodec~\cite{ai2024apcodec} and MDCTCodec~\cite{jiang2024mdctcodec} perform analysis and reconstruction on spectral representations derived from the short-time Fourier transform (STFT) and the modified discrete cosine transform (MDCT), respectively, and both reconstruct the waveform via their corresponding inverse transforms, with RVQ adopted for quantization.
In contrast, Spectral Codec~\cite{langman2024spectral} compresses mel-spectrograms into discrete tokens using multiple finite scalar quantization (FSQ) and reconstructs speech with a neural vocoder-style decoder. 
A key advantage of spectral-domain codecs is that the features are already temporally subsampled by the analysis hop, so the encoder-decoder needs only mild additional downsampling/upsampling, reducing computation and improving deployment efficiency, especially on CPU-only devices. 
Both the aforementioned waveform-domain and spectral-domain codecs typically rely on multiple VQ or FSQ units and operate at moderate bitrates (e.g., 2 kbps), which still fall short of the ultra-low-bitrate regime.


Adopting a single-codebook quantizer offers a direct route to further bitrate reduction; however, excessively aggressive compression can incur severe information loss, rendering the decoded speech markedly unnatural and potentially unintelligible. 
BigCodec~\cite{xin2024bigcodec} employs a single VQ codebook and couples it with a large-capacity encoder–decoder (159M parameters) to maintain speech quality, thereby enabling stable operation in the low-bitrate regime (1.04 kbps for 16-kHz speech coding~\cite{xin2024bigcodec}). 
WavTokenizer~\cite{jiwavtokenizer} instead combines single-layer VQ with aggressive temporal downsampling, and stabilizes codebook usage via k-means initialization and random reactivation, while still relying on a large-capacity encoder–decoder to achieve high-quality low-bitrate compression (0.9 kbps for 24-kHz speech coding~\cite{jiwavtokenizer}). 
{TAAE \cite{parkerscaling} and TS3-Codec \cite{wu2025ts3} also showed that further increasing model capacity with a large-parameter-count Transformer architecture can push speech coding to low bitrates while maintaining strong reconstruction quality.}
However, achieving high-quality ultra-low-bitrate speech coding, under practical constraints on model size and computational cost remains highly challenging.

{Recent advances have also seen the emergence of several neural codecs built upon self-supervised learning (SSL) speech representations. 
For example, FocalCodec \cite{dellafocalcodec} reports strong 16-kHz speech coding performance at ultra low bitrates by leveraging powerful pretrained speech representations together with a low-bitrate codec design. 
SemantiCodec \cite{liu2024semanticodec} also pushes coding toward the ultra-low-bitrate regime by exploiting high-level semantic representations and generative reconstruction. 
However, such approaches rely heavily on pretrained upstream SSL models that are typically designed for low-sampling-rate speech input, which limits their direct applicability to higher-sampling-rate speech coding (e.g., 48 kHz).
As a result, achieving high-quality ultra-low-bitrate speech coding under practical model and computational constraints, while also supporting higher-sampling-rate speech reconstruction, remains highly challenging.}


To address the above issues, we propose FMelCodec, an ultra-low-bitrate spectral-domain speech codec that enables high-quality and stable coding at as low as 250 bps. 
The FMelCodec follows two design principles for bitrate reduction. 
First, it codes mel-spectrograms, which are inherently temporally subsampled for aggressive compression, perceptually structured, and readily convertible to waveforms via mature vocoding. 
Second, it uses single-codebook VQ for aggressive compression and a lightweight conditional flow matching (CFM)~\cite{lipman2023flow} refiner to correct quantization-induced mel distortions, i.e., to enhance the decoding capability. 
{Inspired by \cite{pia2025flowmac}, the proposed FMelCodec adopts a three-stage coding--refinement--reconstruction (CRR) framework formulation for ultra-low-bitrate mel-spectrogram-based speech coding.} 
Within the CRR framework, the mel-spectrogram coding stage performs aggressive compression via single-codebook VQ and decodes a coarse mel-spectrogram. 
Then, the mel-spectrogram refinement stage restores fine-grained mel structures from codec-degraded features to counteract the coarseness introduced by single-codebook quantization at ultra-low bitrates. 
Finally, the vocoding-driven waveform reconstruction stage synthesizes the waveform from the refined mel-spectrogram using a pretrained neural vocoder.
Both objective and subjective evaluations confirm that FMelCodec delivers strong performance at ultra-low bitrates (250 bps for 16-kHz speech and 750 bps for 48-kHz speech), outperforming strong baselines that require substantially higher model and computational complexity.

{Compared with existing speech coding methods, the main novelties and contributions of the proposed FMelCodec are summarized as follows.} 

\begin{itemize}

    \item {FMelCodec pushes speech coding toward a substantially lower bitrate standard, reaching as low as 250 bps, and moves speech codecs toward practical deployment in bandwidth-constrained and low-resource scenarios.}

    \item {FMelCodec adopts a fully acoustic-level CRR framework in the mel-spectrogram domain, incorporating single-codebook discretization, CFM-based refinement, and vocoding-driven reconstruction, while avoiding reliance on SSL-based semantic features and maintaining acceptable complexity.
    }

    
    \item {FMelCodec introduces several effective technical designs, e.g., online-clustering-based single-codebook VQ discretization for bitrate reduction while avoiding codebook collapse, and self-consistency loss to improve the efficiency of few-step CFM refinement.}

\end{itemize}




The rest of this paper is organized as follows. 
Section~\ref{sec:related} reviews prior work on spectral-domain neural speech codecs, CFM applications, and robust neural vocoding technology, which align with the three stages of our CRR framework in FMelCodec, respectively.
Section~\ref{sec:method} provides details on our proposed FMelCodec. 
Section~\ref{sec:Experiments} presents our
experimental results.
Finally, Section~\ref{sec:conclusion} gives conclusions and discusses potential directions for future work.

\section{Related Work}
\label{sec:related}
\subsection{Spectral-domain Neural Speech Codecs}
\label{subsec:spectrum_based_codecs}

Spectral-domain neural speech codecs are particularly well suited to ultra-low-bitrate coding compared with waveform-domain ones, as the waveform-to-spectrum transform inherently performs temporal subsampling, enabling low bitrates without relying on deep stacks of downsampling/upsampling layers. 
Existing spectral-domain codecs can be broadly grouped into three categories according to the spectral representation they use.
\begin{itemize}
\item\textbf{{STFT-Spectrum-based Neural Speech Codecs.}}
\label{subsubsec:stft_spectrum_codecs}
APCodec~\cite{ai2024apcodec} is a representative example of this line of work, performing speech coding on STFT-derived amplitude and phase spectra.
Given an input waveform, it first computes frame-level amplitude and phase features, then encodes them using two dedicated neural branches and fuses the resulting representations in the latent space. 
The fused features are subsequently discretized by RVQ and decoded to reconstruct amplitude and phase spectra, after which an inverse STFT is applied to recover the time-domain waveform. 
Nevertheless, APCodec still relies on RVQ to attain sufficient representational capacity, which places it in a moderate-bitrate regime relative to the ultra-low-bitrate target pursued in this work.

\item\textbf{{MDCT-Spectrum-based Neural Speech Codecs.}}
\label{subsubsec:mdct_spectrum_codecs}
Relative to STFT spectra, MDCT spectra are more compact and real-valued while remaining fully invertible, offering a more favorable target for lightweight modeling. 
Therefore, MDCTCodec~\cite{jiang2024mdctcodec} was proposed to further simplify spectral coding by modeling the MDCT spectrum, using RVQ for discretization and enabling a single encoder--decoder backbone in place of the dual-branch amplitude--phase design.
StreamCodec~\cite{jiang2025streamable} further extends this MDCT-domain perspective to streaming low-latency speech communication scenarios by adopting a fully causal architecture and introducing a novel residual scalar–vector quantization (RSVQ) strategy. 
While MDCT-based approaches can markedly reduce model and computational complexity relative to STFT-based methods, their reconstruction quality still hinges on residual quantization structures, which makes further bitrate reduction difficult.


\item\textbf{{Mel-Spectrogram-based Codecs.}}
\label{subsubsec:mel_spectrum_codecs}
This line of work discretizes speech mel-spectrograms and reconstructs speech via neural vocoding. 
For example, Spectral Codec~\cite{langman2024spectral} first encodes the mel-spectrogram with a residual encoder, then discretizes it using eight parallel FSQ quantizers, and finally concatenates the quantized outputs and reconstructs speech with a HiFi-GAN vocoder~\cite{kong2020hifi}. 
However, the use of parallel FSQ leads to a relatively high bitrate; as reported in~\cite{langman2024spectral}, it operates at 6.9 kbps.
Benefiting from the perceptually aligned structure of mel-spectrograms and mature vocoding techniques for reconstruction, our FMelCodec is also a mel-spectrogram-based method, but it can operate in the ultra-low-bitrate regime (i.e., 250 bps), by combining single-VQ discretization and CFM-based refinement in the mel domain. 
Moreover, FMelCodec explicitly encodes and decodes mel-spectrograms, which eases the decoding burden, whereas Spectral Codec reconstructs waveforms directly from quantized features without an explicit mel decoding (i.e., only performing mel encoding).


\end{itemize}

\subsection{{Diffusion and Conditional Flow Matching for Speech Generation}}
\label{subsec:cfm_related}

{Diffusion models~\cite{ho2020denoising} are a class of continuous-time generative models that construct data generation as the reverse of a gradual noising process. 
Instead of producing the target in a single step, they start from a simple prior such as Gaussian noise and iteratively refine the sample through multiple denoising steps, thereby building a generation trajectory toward the target data distribution. 
Owing to this strong generative capability, diffusion models have been widely applied to a range of speech generation tasks.
In addition to these speech generation tasks, diffusion models have also been introduced into neural speech codecs to recover details lost after aggressive quantization or tokenization. 
For example, LaDiffCodec~\cite{yang2024generative} employs latent diffusion for de-quantization, while MBD~\cite{san2023discrete} reconstructs high-fidelity waveforms from discrete acoustic tokens using multi-band diffusion. 
However, despite their effectiveness, diffusion models usually require many iterative denoising steps during inference, which leads to relatively high computational cost.}

{Compared with diffusion models}, CFM~\cite{lipman2023flow} has recently emerged as an effective framework for conditional generative modeling.
Starting from a simple prior (often Gaussian noise), CFM learns a time-dependent velocity field that transports samples toward the data distribution. 
At inference, samples are obtained by numerically integrating the corresponding ordinary differential equation (ODE) driven by the learned velocity field.


CFM has shown strong potential in various speech generation tasks. 
For example, in text-to-speech (TTS) generation, CFM has been used to transport latent variables into high-quality mel-spectrograms conditioned on linguistic features, enabling fast and natural synthesis~\cite{mehta2024matcha}. 
In speech enhancement, CFM-based formulations learn deterministic mappings from corrupted acoustic features to clean ones, effectively removing noise and artifacts~\cite{wang2025flowse}. 
In speech coding, CFM is also emerging as a promising tool. 
For example, FlowDec~\cite{welkerflowdec} equips a non-adversarial codec with a CFM-based postfilter to enhance the STFT representations of the coded speech, substantially improving reconstruction fidelity by recovering details lost during quantization. 
{In addition, FlowMAC~\cite{pia2025flowmac} explores CFM for low-bitrate coding by incorporating a CFM-based generative reconstruction module, which improves decoded mel-spectrogram quality from compressed representations. 
However, FlowDec and FlowMAC still need  to operate at a moderate bitrate, i.e., 4 kbps for 48-kHz speech coding, as reported in~\cite{welkerflowdec} and 3 kbps for 24-kHz speech coding as reported in \cite{pia2025flowmac}. 
Beyond speech coding, MuCodec~\cite{xu2024mucodec} further extends flow-matching-based generative reconstruction to the task of ultra-low-bitrate music coding.
}

Building on the demonstrated effectiveness of CFM in speech generation, we integrate it into FMelCodec as a dedicated refinement module that enhances codec-degraded mel-spectrograms under ultra-low bitrates before neural vocoding. 
Unlike prior work that performs refinement in the STFT domain~\cite{welkerflowdec}, we refine directly in the mel domain, which reduces the difficulty of the enhancement problem. 
We further introduce a self-consistency training scheme that enables fewer ODE solver steps at inference, reducing computational cost.


\subsection{Neural Vocoding for Robust Waveform Reconstruction}
\label{subsec:vocoding_related}

Neural vocoding has demonstrated strong capability in converting mel-spectrograms into high-fidelity waveforms, and importantly, they remain effective even when the input mel features are synthetic or moderately perturbed. 
This robustness has been repeatedly validated in modern TTS pipelines, where upstream models predict ``imperfect'' mel-spectrograms and a neural vocoder still produces natural-sounding speech~
\cite{kong2020hifi}. 
Such a property motivates separating acoustic representation modeling from waveform reconstruction. 
In our proposed FMelCodec, we first compress and refine mel-spectrograms, and then rely on a pretrained vocoder to recover time-domain details, thereby simplifying codec training while retaining high-quality reconstruction.

\begin{figure*}
    \centering
    \includegraphics[width=1\linewidth]{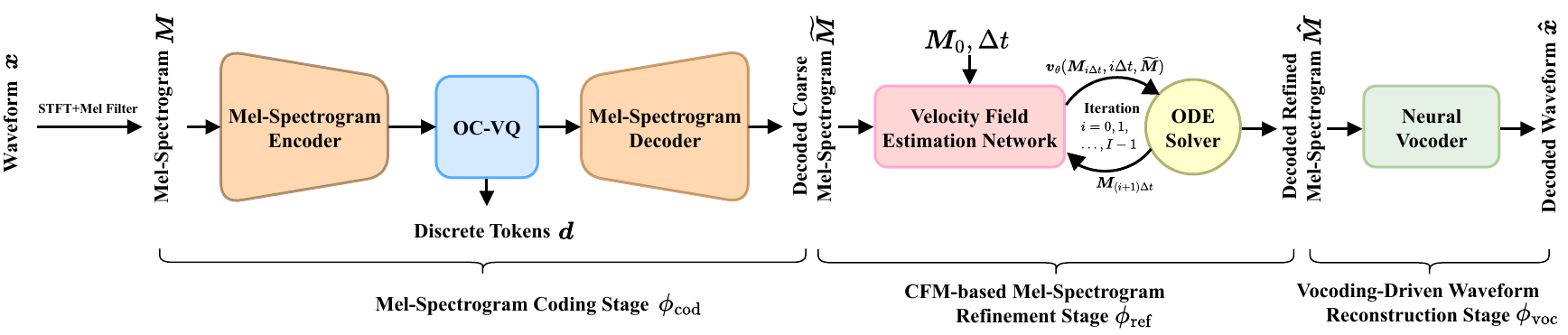}
\caption{Inference pipeline of the proposed FMelCodec under the CRR framework.}
    \label{fig:1}
\end{figure*}

\begin{figure*}
    \centering
    \includegraphics[width=\linewidth]{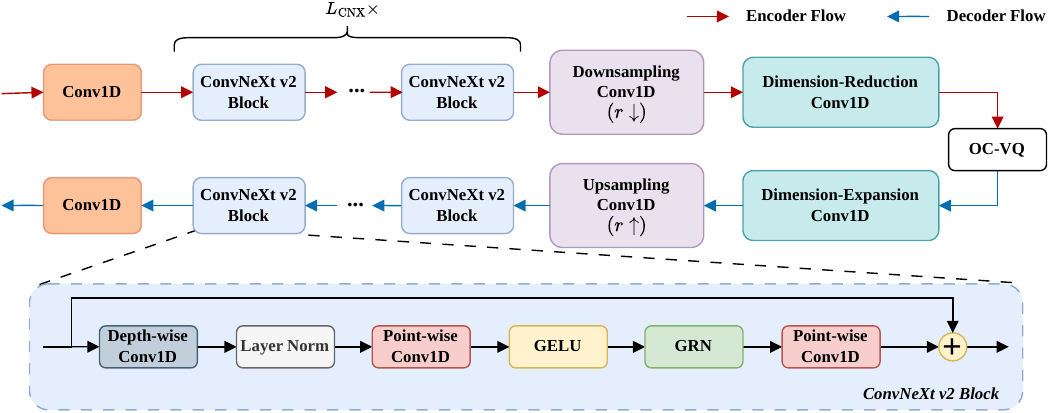}
\caption{{Architecture of the ConvNeXt v2--based mel-spectrogram encoder and decoder used in the mel-spectrogram coding stage $\phi_{\text{cod}}$.}}

    \label{fig:conv}
\end{figure*}

\section{Proposed Method}
\label{sec:method}

An overview of our proposed ultra-low-bitrate speech codec, FMelCodec, is shown in Fig.~\ref{fig:1}. 
It adopts a three-stage CRR framework, consisting of a mel-spectrogram coding stage, a CFM-based mel-spectrogram refinement stage, and a vocoding-driven waveform reconstruction stage. 
The three stages are trained separately and are connected sequentially during inference.
Overall, FMelCodec aggressively compresses the input waveform $\bm{x}\in\mathbb{R}^{T}$ into discrete tokens $\bm{d}=[d_1,d_2,\dots,d_{N'}]^\top\in\mathbb N^{N'}$ and reconstructs high-quality speech $\hat{\bm{x}}\in\mathbb{R}^{T}$, with all operations performed in the mel-spectrogram domain.
Here, $T$ denotes the waveform length, and $R=\frac{T}{N'}$ denotes the downsampling factor from the waveform to discrete tokens. 
Next, we describe the three stages in the CRR framework adopted by FMelCodec in detail, including their respective roles, model architectures, and training procedures.


\subsection{Mel-Spectrogram Coding}
\label{subsec:coding_stage}

The mel-spectrogram coding stage $\phi_{\text{cod}}$ compresses the mel-spectrogram $\bm{M}\in\mathbb{R}^{N\times D}$ extracted from $\bm{x}$ into a compact discrete token sequence $\bm{d}$ and produces a decoded coarse mel-spectrogram $\widetilde{\bm{M}}\in\mathbb{R}^{N\times D}$ to serve as input for the subsequent refinement stage, where $N$ and $D$ denote the number of frames and mel-frequency bins of the mel-spectrogram, respectively, i.e., 
\begin{equation}
    \bm{d},\,\widetilde{\bm{M}}=\phi_{\text{cod}}(\bm{M}).
\end{equation}
Specifically, as shown in Fig. \ref{fig:1}, the mel-spectrogram coding stage $\phi_{\text{cod}}$ consists of a mel-spectrogram encoder, an OC-VQ, and a mel-spectrogram decoder.


\subsubsection{\textbf{Mel-Spectrogram Encoder \& Decoder}}
\label{subsubsec:mel_encdec}

As shown in Fig.~\ref{fig:conv}, we build the mel-spectrogram encoder and decoder on ConvNeXt v2~\cite{woo2023convnext}. 
ConvNeXt-style backbones have been adopted for mel-spectrogram modeling in speech generation and vocoding pipelines, demonstrating strong empirical performance~\cite{lv2024freev,du2023apnet2,siuzdak2023vocos}.
A ConvNeXt v2 block is a residual unit that combines large-kernel depth-wise convolution for local temporal aggregation with lightweight channel mixing. 
Concretely, the input is first processed by a depth-wise convolution to capture local context, then normalized, and passed through a point-wise convolution that expands the channel dimension. 
After a Gaussian error linear unit (GELU) 
nonlinearity, a global response normalization (GRN) layer modulates channel responses, and a second point-wise convolution projects the features back to the original dimensionality. Finally, a residual connection adds the block input to the block output, which stabilizes optimization while preserving information flow.

The mel-spectrogram encoder and decoder are designed symmetrically around this backbone. 
We treat the input mel-spectrogram $\bm{M}\in\mathbb{R}^{N\times D}$ as a length-$N$ temporal sequence, where the $D$ mel bins are treated as channels. 
The mel-spectrogram encoder first applies an input convolutional layer to project the $D$-channel mel features to a hidden width, followed by $L_{CNX}$ stacked ConvNeXt v2 blocks to model temporal structure. To further reduce the bitrate, we append a downsampling convolutional layer after the ConvNeXt v2 backbone to downsample the temporal resolution by a factor of $r$ (from $N$ to $N'$, i.e. $r=\frac{N}{N'}$), and then apply a dimension-reduction convolutional layer to reduce the channel dimension for subsequent quantization, producing the encoded latent feature $\bm{Z}=[\bm{z}_1,\ldots,\bm{z}_{N'}]^\top\in\mathbb{R}^{N'\times C}$, where $C$ is the feature dimension. 

The mel-spectrogram decoder takes $\hat{\bm{Z}}=[\hat{\bm{z}}_1,\ldots,\hat{\bm{z}}_{N'}]^\top\in\mathbb{R}^{N'\times C}$ as input and mirrors the mel-spectrogram encoder in reverse: it first applies a channel-expansion convolution, then an upsampling convolution (with upsampling factor $r$) to restore the number of frames to $N$, and finally refines the upsampled features with another stack of $L_{CNX}$ ConvNeXt v2 blocks.
A final output convolutional layer projects features back to the mel domain, producing the decoded coarse mel-spectrogram $\widetilde{\bm{M}}\in\mathbb{R}^{N\times D}$, which serves as the conditioning input to the subsequent CFM-based mel-spectrogram refinement stage.

\subsubsection{\textbf{Single-Codebook Vector Quantizer with Online Clustering}}
\label{subsubsec:online_clustering}

We adopt a single-codebook VQ to discretize the encoder outputs for ultra-low-bitrate compression. 
The VQ has a trainable codebook, denoted as $\mathbb W=\{\bm{w}_k\in\mathbb{R}^{C}\mid k=1,\ldots,K\}$,
where $K$ is the codebook size. 
Given the $n$-th frame encoded latent feature $\bm{z}_n$ in $\bm{Z}$, its discrete token $d_n$ and quantized result $\hat{\bm{z}}_n$ in $\hat{\bm{Z}}$ are determined based on the minimum Euclidean distance between $\bm{z}_n$ and the codevectors in $\mathbb{W}$, as expressed by the following formulas:
\begin{equation}
    d_n = \arg\min_{k \in \{1,2,\dots,K \}} \| \bm{z}_n - \bm{w}_k \|_2,
\end{equation}
\begin{equation}
    \hat{\bm{z}}_n = \bm{w}_{d_n},
\end{equation}
where $n=1,2,\dots,N'$. 
The tokens $\bm{d} = [d_1, d_2, \dots, d_{N'}]^\top$ are the discrete representation of the speech $\bm{x}$, which can be used for transmission, storage, and other purposes.
Let $f_s$ be the waveform sampling rate, $w_s$ be the frame shift when extracting the mel-spectrogram, then the bitrate of the tokens is
\begin{equation}
    \mathrm{Bitrate}
    =
    \frac{f_s}{r\,w_s}\,\log_2 K
    \quad \text{(bps)},
    \label{eq:bitrate_singlecodebook}
\end{equation}
where $r\,w_s$ is the downsampling factor from the waveform to token, i.e., $r\,w_s=R$. 

However, a naive single-codebook VQ is prone to \emph{codebook collapse}, where a small subset of codevectors dominate the assignments and many entries are rarely selected~\cite{baevski2020wav2vec,zheng2023online,zheng2025ervq}.
Codebook collapse leads to inefficient utilization of the codebook, reducing its representational capacity, limiting the effectiveness of the quantization process, and ultimately degrading the quality of the decoded mel-spectrogram.
To overcome this issue, we incorporate an online clustering mechanism inspired by \cite{zheng2025ervq} to construct OC-VQ, which monitors codevector activity during training and dynamically relocates underused codevectors to better align with the evolving feature distribution.

\begin{figure}
    \centering
    \includegraphics[width=1\linewidth]{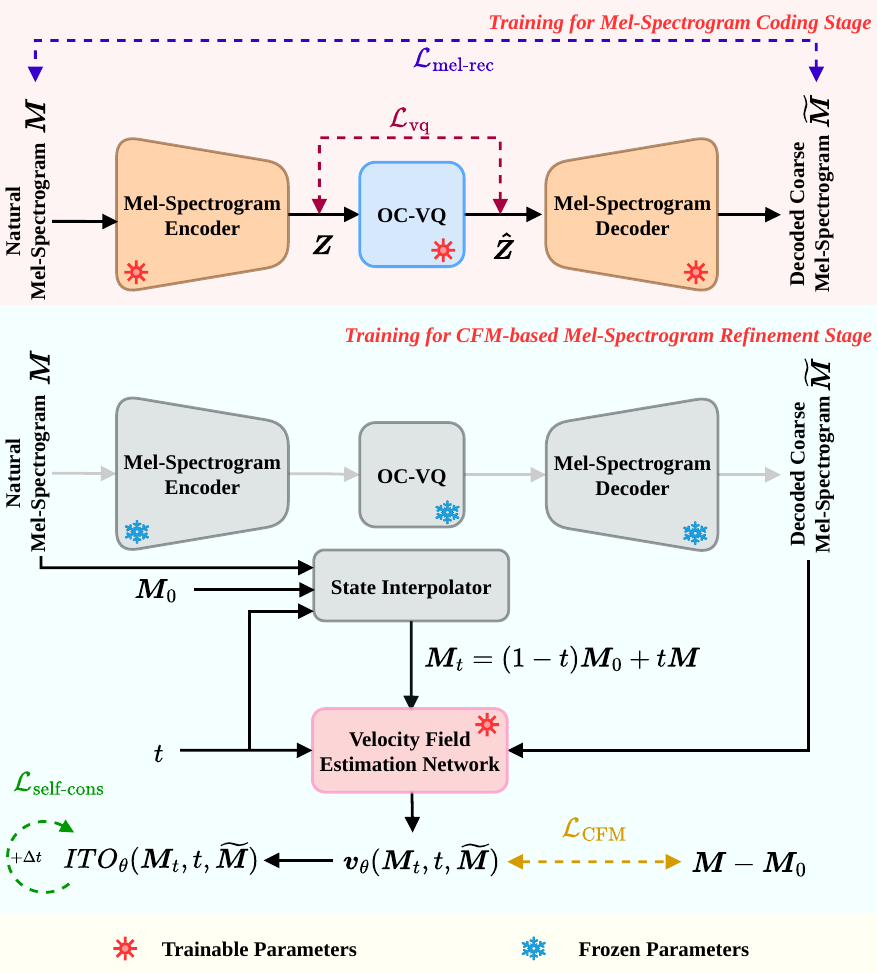}
\caption{Training process of the mel-spectrogram coding stage $\phi_{\text{cod}}$ and CFM-based mel-spectrogram refinement stage $\phi_{\text{ref}}$. {The state interpolator computes the intermediate state $\bm{M}_t=(1-t)\bm{M}_0+t\bm{M}$ and contains no trainable parameters.}}
    \label{fig:placeholder}
\end{figure}

The online clustering mechanism is employed solely during the training of OC-VQ, without imposing additional computational overhead or affecting the inference process.
Specifically, at the $\tau$-th training iteration, we form a batch of $N^{(batch)}$ encoded feature vectors $\{\bm{z}_n^{(\tau)}\}_{n=1}^{N^{(batch)}}$ to be quantized.
Let $c_k^{(\tau)}$ denote the number of times the $k$-th codevector $\bm{w}_k$ is assigned in the mini-batch at $\tau$-th training iteration. 
We maintain an exponential moving average (EMA) estimate of the usage rate, denoted by $\pi_k^{(\tau)}$, as
\begin{equation}
\pi_k^{(\tau)}=\rho\,\pi_k^{(\tau-1)}+(1-\rho)\,\frac{c_k^{(\tau)}}{N^{(batch)}},
\label{eq:ema_usage}
\end{equation}
where $\rho\in(0,1)$ is the EMA factor and $\pi_k^{(\tau-1)}$ is the usage rate of the previous (i.e., $(\tau-1)$-th) training iteration. 
Based on $\pi_k^{(\tau)}$, we compute a data-dependent refresh coefficient as
\begin{equation}
\gamma_k^{(\tau)}
=\exp\!\left(-\frac{10\, \pi_k^{(\tau)}\,K}{1-\rho}-\delta\right),
\label{eq:decay_coeff}
\end{equation}
where $\delta$ is a small offset. 
This design yields larger $\gamma_k^{(\tau)}$ for rarely used codevectors, while keeping frequently used ones relatively stable. 
The $k$-th codevector $\bm{w}_k^{(\tau)}$ at $\tau$-th training iteration is then updated by
\begin{equation}
\bm{w}_k^{(\tau)}
=\bigl(1-\gamma_k^{(\tau)}\bigr)\bm{w}_k^{(\tau-1)}
+\gamma_k^{(\tau)}\,\bm{a}_k^{(\tau)},
\label{eq:codeword_update}
\end{equation}
where $\bm{w}_k^{(\tau-1)}$ is the codevector from the previous (i.e., $(\tau-1)$-th) training iteration.
Vector $\bm{a}_k^{(\tau)}$ is an anchor feature sampled from the current mini-batch using a probabilistic random anchor sampling strategy. 
Specifically, given the set of encoded feature vectors $\{\bm{z}_n^{(\tau)}\}_{n=1}^{N^{(batch)}}$ in the $\tau$-th training iteration, we compute their Euclidean distances to the current $(k-1)$-th codevector $\bm{w}_k^{(\tau-1)}$ as $\Delta_{n,k}^{(\tau)}=\|\bm{z}_n^{(\tau)}-\bm{w}_k^{(\tau-1)}\|_2, n=1,2,\dots,N^{(batch)}$. 
A categorical distribution over the $N^{(batch)}$ candidates is then formed by softmax-normalizing $\{\Delta_{n,k}^{(\tau)}\}_{n=1}^{N^{(batch)}}$, and one vector is sampled as the anchor $\bm{a}_k^{(\tau)}$ according to this distribution. 
This distance-based sampling tends to relocate underutilized codevectors toward regions that are not well covered by the current codebook, while the refresh coefficient $\gamma_k^{(\tau)}$ controls the update magnitude so that frequently used codevectors remain stable. 

\subsubsection{\textbf{Reconstruction-based Training Criteria}}
\label{subsubsec:codec_training}

As shown in Fig. \ref{fig:placeholder}, unlike most existing adversarial-based neural codecs \cite{langman2024spectral,li2024single}, the mel-spectrogram coding stage $\phi_{cod}$ of FMelCodec is trained in a purely reconstruction-based, non-adversarial manner.


We first define the mel-spectrogram reconstruction loss $\mathcal{L}_{\text{mel-rec}}$ as a weighted combination of $\ell_1$ and $\ell_2$ errors to encourage the decoded coarse $\widetilde{\bm{M}}$ to match the natural $\bm{M}$ at the time--frequency bin level, i.e.,
\begin{equation}
\mathcal{L}_{\text{mel-rec}}
=
\bigl\|\bm{M}-\widetilde{\bm{M}}\bigr\|_1
+
\bigl\|\bm{M}-\widetilde{\bm{M}}\bigr\|_2^2.
\label{eq:rec_loss}
\end{equation}

To regularize vector quantization under the single-codebook setting, we then introduce a VQ term ${\mathcal{L}_{\text{vq}}}$ {based on the standard vector quantized variational autoencoder (VQ-VAE) formulation}, which accounts for both the codebook update and encoder commitment effects~\cite{van2017neural,razavi2019generating}.
The OC-VQ regularization term is written as
\begin{equation}
{\mathcal{L}_{\text{vq}}}
=
\bigl\|\mathrm{sg}[\bm{Z}]-\hat{\bm{Z}}\bigr\|_2^2
+
\eta\,\bigl\|\bm{Z}-\mathrm{sg}[\hat{\bm{Z}}]\bigr\|_2^2,
\label{eq:ocvq_loss}
\end{equation}
where $\mathrm{sg}[\cdot]$ is the stop-gradient operator and $\eta$ controls the commitment strength.

Finally, the overall training objective of the mel-spectrogram coding stage is defined as follows, to jointly train the mel-spectrogram encoder, OC-VQ, and mel-spectrogram decoder:
\begin{equation}
\mathcal{L}_{\text{cod}}
=
\lambda_{\text{mel-rec}}\,\mathcal{L}_{\text{mel-rec}}
+
{\lambda_{\text{vq}}\,\mathcal{L}_{\text{vq}}},
\label{eq:cod_loss}
\end{equation}
where $\lambda_{\text{mel-rec}}$ and ${\lambda_{\text{vq}}}$ are scalar hyperparameters controlling the relative weights.

\subsection{Conditional Flow Matching based Mel-Spectrogram Refinement}
\label{subsec:refinement_stage}

In the preceding stage, despite the incorporation of reconstruction losses, the quantization mechanism remains relatively simplistic, inevitably resulting in a decoded coarse mel-spectrogram $\widetilde{\bm{M}}$. 
To address this limitation, the CFM-based mel-spectrogram refinement stage, denoted as $\phi_{\text{ref}}$, is introduced to further refine $\widetilde{\bm{M}}$, thereby enhancing the overall mel-spectrogram quality and yielding a more detailed and high-fidelity representation $\hat{\bm{M}}\in\mathbb{R}^{N\times D}$. 
Specifically, this stage learns a conditional generative transformation based on CFM~\cite{lipman2023flow}, which derives a refined mel-spectrogram $\hat{\bm{M}}$ from an initial distribution $\bm{M}_0\sim \mathcal P_0$, conditioned on the coarse mel-spectrogram $\widetilde{\bm{M}}$, where $\mathcal P_0=\mathcal{N}\!\left(\mathbf{0},\, \bm{I}_{N\times D}\right)$ is a Gaussian distribution and $\bm{I}_{N\times D}$ denotes the identity covariance matrix in the $N \times D$-dimensional space, i.e., 
\begin{equation}
    \hat{\bm{M}}=\phi_{\text{ref}}(\bm{M}_0|\widetilde{\bm{M}}).
\end{equation}



\begin{figure}
    \centering
    \includegraphics[width=1\linewidth]{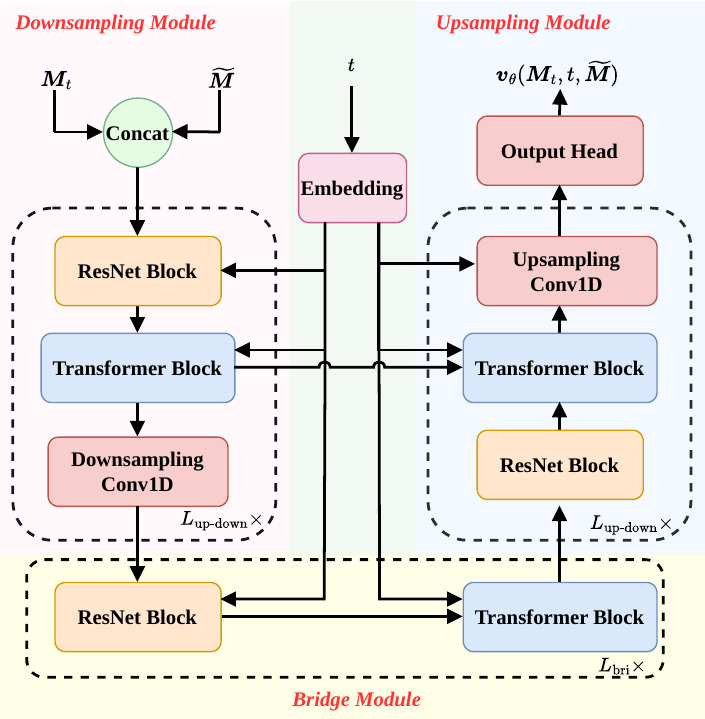}
    \caption{{Architecture of the velocity field estimation network $\theta$ used in the CFM-based mel-spectrogram refinement stage $\phi_{\text{ref}}$.}}
    \label{fig:cfm_net}
\end{figure}

\subsubsection{\textbf{Problem Definition and ODE-based Solution}}
\label{subsubsec:refinement_stage}
The CFM defines a time-dependent state $\bm{M}_t$ in the mel feature space and specifies the corresponding velocity field $\bm{v}_\theta(\bm{M}_t, t, \widetilde{\bm{M}})$ at time $t$, conditioned on $\widetilde{\bm{M}}$. 
These quantities follow a differential relationship given by
\begin{equation}
\frac{\mathrm{d}\bm{M}_t}{\mathrm{d}t}
=
\bm{v}_\theta(\bm{M}_t,\, t,\, \widetilde{\bm{M}}),\quad t\in[0,1].
\end{equation}
The velocity field $\bm{v}_\theta$ is parameterized by a neural network with parameters $\theta$. 
The refined mel-spectrogram $\hat{\bm{M}}$ (i.e., the terminal state) can be obtained from the initial distribution $\bm{M}_0$ by solving the following integral equation:
\begin{equation}
\label{eq:int}
    \hat{\bm{M}}=\bm{M}_0+\int_0^1 \bm{v}_\theta(\bm{M}_t,t,\widetilde{\bm{M}})\,\mathrm{d}t.
\end{equation}


To efficiently obtain $\hat{\bm{M}}$, we employ an ODE-based solver at inference time. 
Specifically, given $\widetilde{\bm{M}}$, we draw an initial sample $\bm{M}_0$ and approximate the integral in~\eqref{eq:int} by numerically solving the ODE from $t=0$ to $t=1$. 
Starting from $\bm{M}_0$, the solution is iteratively updated with a step size of $\Delta t = 1 / I$ using an explicit Euler scheme, thereby estimating the integral in~\eqref{eq:int}, i.e., 
\begin{equation}
    \bm{M}_{(i+1)\Delta t}=\bm{M}_{i\Delta t}+\Delta t\cdot\,\bm{v}_\theta(\bm{M}_{i\Delta t},i\Delta t,\widetilde{\bm{M}}),
\end{equation}
where $i=0,\ldots,I-1$, and $I$ denotes the number of iterations. 
Finally, we take the final state, i.e., $\bm{M}_{I\Delta t}$, as the decoded refined mel-spectrogram $\hat{\bm{M}}$.

\subsubsection{\textbf{Velocity Field Estimation Network}}

To estimate the velocity field $\bm{v}_\theta(\bm{M}_t, t, \widetilde{\bm{M}})$, we adopt a neural network $\theta$ as illustrated in Fig.~\ref{fig:cfm_net}. 
Given the current state $\bm{M}_t$ and conditioning $\widetilde{\bm{M}}$, we concatenate them along the channel dimension to form the input feature map. 
The time variable $t \in [0,1]$ is also embedded and incorporated into the network, enabling the joint estimation of the velocity field $\bm{v}_\theta(\bm{M}_t, t, \widetilde{\bm{M}})$ at the current time step.

Specifically, the velocity field estimation network $\theta$ adopts a {TransformerUNet} architecture {used in}~\cite{mehta2024matcha}, consisting of a downsampling module, a bridge module, and an upsampling module arranged in an encoder–decoder fashion. 
The downsampling module is composed of $L_{\text{up-down}}$ downsampling submodules, each operating at a different resolution level. 
For each downsampling submodule, the network first applies a convolutional ResNet block \cite{he2016identity} followed by a Transformer block \cite{vaswani2017attention}, and then reduces the temporal resolution using a strided 1D convolution. 
The intermediate feature maps produced by the Transformer block are cached and later used as skip connections to the corresponding stages of the upsampling module.
The bridge module operates at a shared lowest resolution and consists of $L_{bri}$ middle blocks, each comprising a ResNet block followed by a Transformer block, to enable deeper feature processing before upsampling.
The upsampling module mirrors the downsample hierarchy and is composed of $L_{\text{up-down}}$ upsampling submodules. 
For each submodule, it first upsamples the temporal resolution using a transposed 1D convolution, then concatenates the corresponding skip feature from the downsampling module with the Transformer block input, applies a Transformer block, and finally processes the features with a ResNet block. 
After the final upsampling stage, a $D$-dimensional output head projects the decoded features to the predicted velocity. 
To better model fine-grained periodic structures in speech, we employ feed-forward layers with SnakeBeta activations~\cite{lee2023bigvgan} in all Transformer blocks.
For the time variable $t\in[0,1]$, it is embedded using a sinusoidal embedding followed by a multi-layer perceptron (MLP)~\cite{popov2021grad} and injected into all ResNet and Transformer blocks as timestep conditioning, where the same timestep embedding is shared across all temporal positions.
\subsubsection{\textbf{Self-Consistency Training Criteria}}
\label{subsec:self_consistency_cfm}
As illustrated in Fig.~\ref{fig:placeholder}, the training of the CFM-based mel-spectrogram refinement stage is conducted on top of a pre-trained mel-spectrogram coding stage. 
The mel-spectrogram coding stage provides decoded coarse mel-spectrogram $\widetilde{\bm{M}}$ for refinement, while the natural mel-spectrogram $\bm{M}$ is used as the training target.

To facilitate efficient training, we adopt the optimal-transport CFM (OT-CFM) construction, in which a linear path is defined between the prior sample $\bm{M}_0$ and the target mel-spectrogram $\bm{M}$, i.e., 
\begin{equation}
\label{equ15}
 \bm{M}_t=(1-t)\bm{M}_0+t\bm{M}, \quad t\in[0,1].   
\end{equation}
Along this path, the differential of $\bm{M}_t$ is given by
\begin{equation}
    \frac{\mathrm{d}\bm{M}_t}{\mathrm{d}t}=\bm{M}-\bm{M}_0,
\end{equation}
which does not depend on $t$. 
A straightforward and commonly used training strategy for CFM is to force the output of the velocity-field estimation network $\bm{v}_\theta(\bm{M}_t, t, \widetilde{\bm{M}})$ to approximate the target $\bm{M}-\bm{M}_0$, i.e.,
\begin{equation}
\mathcal{L}_{\text{CFM}}
=
\mathbb{E}_{\bm{M}_0\sim \mathcal{P}_0,\,t\sim\mathcal{U}(0,1)}
\Bigl\|
\bm{v}_\theta(\bm{M}_t,t,\widetilde{\bm{M}})-(\bm{M}-\bm{M}_0)
\Bigr\|_2^2.
\label{eq:cfm_loss}
\end{equation}
Although the objective of $\mathcal{L}_{\text{CFM}}$ is to enforce the velocity field to approximate a time-invariant target $\bm{M}-\bm{M}_0$ at any intermediate time, this assumption is idealized. 
In practice, the learned velocity field still exhibits non-negligible time dependence, and satisfactory refinement typically requires solving the ODE with a relatively large number of integration steps (i.e., a large $I$). 
This reliance on multi-step ODE integration consequently increases the computational cost during inference.

To overcome the aforementioned issue, we propose a self-consistency loss that explicitly encourages the refinement dynamics to {remain approximately straight and }time-invariant along the transport path, thereby enabling accurate refinement with fewer ODE integration steps and reducing the overall computational cost. 
Referring to Eq.~\ref{eq:int}, for any time $t$, the intermediate state $\bm{M}_t$ can be evolved into the terminal state by integrating the velocity field. Accordingly, we define a time-dependent integral-form {terminal operator (TO)} form as
\begin{equation}
TO_{\theta}^{\mathrm{int}}(\bm{M}_t,t,\widetilde{\bm{M}})
\triangleq
\bm{M}_t+\int_t^{1}\bm{v}_\theta(\bm{M}_\xi,\xi,\widetilde{\bm{M}})\,\mathrm{d}\xi.
\end{equation}
In the ideal case, the velocity field $\bm{v}_\theta(\bm{m}_\xi,\xi,\widetilde{\bm{M}})$ should be independent of time, as expected in Eq.~\ref{eq:cfm_loss}. 
Under this assumption, the ideal terminal operator (ITO) can be written as follows:
\begin{equation}
ITO_\theta(\bm{M}_t,t,\widetilde{\bm{M}})
\triangleq
\bm{M}_t+(1-t)\,\bm{v}_\theta(\bm{M}_t,t,\widetilde{\bm{M}}).
\label{eq:self_terminal_operator}
\end{equation}
The self-consistency loss is designed to encourage the ITO to produce similar predictions at neighboring time points along the same flow, such that the terminal state remains consistent regardless of the specific intermediate time from which the refinement process is initiated. 
Concretely, we sample $t$ from a truncated Gaussian distribution $\mathcal P_\epsilon =\mathcal{N}_\epsilon (0, \sigma^2)$, restricted to the interval $[0, 1-\epsilon]$,
where $\sigma$ denotes the standard deviation and $\epsilon > 0$ is a small constant introduced to avoid numerical instability near $t = 1$. 
We then sample the time interval $\Delta t$ from a uniform
distribution $\mathcal P_u=\mathcal{U}(\Delta t_{\min}, \Delta t_{\max})$, where
$0 \le \Delta t_{\min} < \Delta t_{\max} {<<} 1$.
We perform a one-step self roll-out via Euler discretization to estimate
$\bm{M}_{t+\Delta t}$ from $\bm{M}_t$, i.e.,
\begin{equation}
\bm{M}_{t+\Delta t}\approx \bm{M}_t+\Delta t\cdot\,\bm{v}_\theta(\bm{M}_t,t,\widetilde{\bm{M}}).
\label{equ20}
\end{equation}
{When $t+\Delta t < 1-\epsilon$, according to the design principle of the self-consistency loss, the ITOs at time steps $t$ and $t+\Delta t$ should be close to each other, i.e., this explicitly enforces $ITO_\theta(\bm{M}_t,t,\widetilde{\bm{M}})=ITO_\theta(\bm{M}_{t+\Delta t}, t+\Delta t, \widetilde{\bm{M}})$. 
By substituting Eqs.~(\ref{equ15}), (\ref{eq:self_terminal_operator}), and (\ref{equ20}) into this expression, we can arrive at an equivalent form that explicitly enforces $\bm{v}_\theta(\bm{M}_t,t,\widetilde{\bm{M}})=\bm{v}_\theta(\bm{M}_{t+\Delta t}, t+\Delta t, \widetilde{\bm{M}})$. 
Therefore, we define the self-consistency loss as}
\begin{equation}
\begin{aligned}
{\mathcal{L}_{\text{self-cons}}
={}} & 
{\mathbb{E}_{t\sim \mathcal{P}_\epsilon,\,\Delta t\sim\mathcal{P}_u}
\Bigl\|
\bm{v}_\theta(\bm{M}_t,t,\widetilde{\bm{M}})}
-\\
&\quad {
\bm{v}_\theta(\bm{M}_{t+\Delta t}, t+\Delta t, \widetilde{\bm{M}})
\Bigr\|_2^2,}
\label{eq:self_consistency_loss}
\end{aligned}
\end{equation}
{under the condition that $t+\Delta t < 1-\epsilon$.
When $t+\Delta t \ge 1-\epsilon$, i.e., with $t$ approaching the terminal time step, we encourage the ITO at time step $t$ to approach the target $\bm{M}$, i.e., explicitly forcing $ITO_\theta(\bm{M}_t,t,\widetilde{\bm{M}})=\bm{M}$. 
By substituting Eqs.~(\ref{equ15}) and (\ref{eq:self_terminal_operator}), this expression is equivalent to explicitly enforcing $\bm{v}_\theta(\bm{M}_t,t,\widetilde{\bm{M}})=\bm{M}-\bm{M}_0$, which plays the same role as $\mathcal{L}_{\text{CFM}}$ in Eq.~(\ref{eq:cfm_loss}). 
Therefore, we set $\mathcal{L}_{\text{self-cons}}=0$ when $t+\Delta t \ge 1-\epsilon$.
}


Finally, in the CFM-based mel-spectrogram refinement stage, the loss function for training the velocity field estimation network $\theta$ is defined as a combination of the CFM loss $\mathcal{L}_{\text{CFM}}$ and the self-consistency loss $\mathcal{L}_{\text{self-cons}}$, i.e.,
\begin{equation}
\mathcal{L}_{\text{ref}}
=
\lambda_{\text{CFM}}\mathcal{L}_{\text{CFM}}
+
\lambda_{\text{self-cons}}\mathcal{L}_{\text{self-cons}},
\label{eq:ref_loss}
\end{equation}
where $\lambda_{\text{CFM}}$ and $\lambda_{\text{self-cons}}$ are scalar hyperparameters controlling the relative weights. 
Under this design, the two losses are combined in a complementary manner. 
$\mathcal{L}_{\text{CFM}}$ encourages the network to learn the correct target dynamics, while $\mathcal{L}_{\text{self-cons}}$ promotes consistency in evolving toward the same terminal state across different time instants. 
As a result, the transformation from the initial distribution to the target distribution can be achieved with fewer ODE integration steps. 


\subsection{Vocoding-Driven Waveform Reconstruction}
\label{subsec:vocoding_stage}

The vocoding-driven waveform reconstruction stage $\phi_{\text{voc}}$ converts the refined mel-spectrogram $\hat{\bm{M}}$ generated in the previous stage into a time-domain speech waveform $\hat{\bm{x}}$ using a neural vocoder, thereby completing the final step of waveform decoding, i.e., 
\begin{equation}
    \hat{\bm{x}}=\phi_{\text{voc}}(\hat{\bm{M}}).
\end{equation}
We choose the HiFi-GAN vocoder \cite{kong2020hifi} because it exhibits strong reconstruction capability and good robustness to moderate deviations in mel-spectrogram inputs, while also maintaining a compact model size. 
Unlike the preceding two stages, the vocoding-driven waveform reconstruction stage is trained in isolation. 
HiFi-GAN is trained directly on natural mel-spectrograms as input, without relying on or waiting for the completion of the other stages.

\section{Experiments and Results}
\label{sec:Experiments}

\subsection{Datasets}

\label{subsec: Data and feature configuration}

We conducted experiments on two speech corpora with different sampling rates, i.e., the LibriTTS~\cite{zen2019libritts} and VCTK~\cite{yamagishi2019cstr} datasets. 
LibriTTS is a multi-speaker English audiobook dataset sampled at 16~kHz; in our experiments, we used the \texttt{train-clean-100} and \texttt{train-clean-360} subsets for training, while \texttt{dev-clean} and \texttt{test-clean} were used for validation and testing, respectively. 
This configuration provides a clean, phonetically rich training set with disjoint speakers for evaluation. 
VCTK is a 48~kHz multi-speaker corpus of read English speech with diverse accents. 
Following the standard split in \cite{jiang2024mdctcodec}, we used 40{,}936 utterances for training and 2{,}937 utterances for testing, enabling us to assess neural speech codecs under a higher sampling rate and a wider range of speaker characteristics.

\subsection{Model Configurations}
\label{subsec:model_config}

In FMelCodec, mel-spectrograms were extracted from speech waveform using an STFT with a frame length of 640 samples, a frame shift of 160 samples (i.e., $w_s=160$), and an FFT size of 1024, together with mel filter of 80 bins for 16 kHz speech (i.e., $D=80$) and 128 bins for 48 kHz speech (i.e., $D=128$). 
The configuration details of the three stages in FMelCodec are as follows.

\begin{itemize}
\item\textbf{Configurations of Mel-Spectrogram Coding Stage.} 
{
For the encoder, the input 1D convolution had a kernel size of 7 and 256 channels. 
The ConvNeXt v2 backbone adopted 8 ConvNeXt v2 blocks (i.e., $L_{CNX}=8$).  
Within each block, the 1D depth-wise convolution used a kernel size of 7, with both the number of convolution groups and the channel size set to 256. 
The first point-wise layer expanded the channel size to 512, whereas the second reduced it back to 256. 
The 1D downsampling convolution after the ConvNeXt v2 backbone had a kernel size of 7, a channel size of 256, and a stride of 4 (i.e., $r=4$).
The final 1D dimension-reduction convolution had a kernel size of 7 and 32 channels, resulting in a 32-dimensional encoded latent feature (i.e., $C=32$). 
The decoder architecture was symmetric to that of the encoder. 
The differing configurations included a channel size of 256 for the 1D dimension-expansion convolution, a kernel size of 16 for the transposed 1D upsampling convolution, and 80 output channels for the final 1D output convolution at 16 kHz and 128 at 48 kHz.
}
For discretization, we employed an OC-VQ with a single codebook of size 1024 (i.e., $K=1024$), using an EMA factor of $\rho=0.999$ and an offset of $\delta=10^{-3}$. 
Under this configuration, FMelCodec achieved an ultra-low bitrate of only 250 bps for 16 kHz speech and 750 bps for 48 kHz speech.
During training, we used 1-second waveform segments with a batch size of 16 for 1M optimization steps (i.e., $N^{(batch)}=400$ for 16 kHz and $N^{(batch)}=1200$ for 48 kHz). 
The training loss hyperparameters were set to $\lambda_{\text{mel-rec}}=45$, ${\lambda_{\text{vq}}}=2.5$, and $\eta=4$. 
The training adopted an AdamW optimizer~\cite{loshchilovdecoupled} with $(\beta_1,\beta_2)=(0.8,0.99)$ and an initial learning rate of $2\times10^{-4}$, which was exponentially decayed by a factor of $0.999$ per epoch. 

\item\textbf{Configurations of CFM-based Mel-Spectrogram Refinement Stage.}
The velocity field estimation network in this stage employed two downsampling submodules, one bridge submodule, and two upsampling submodules (i.e., $L_{up-down}=2$ and $L_{bri}={2}$) {with a hidden channel size of 256 throughout the network. 
In each ResNet block, the main branch comprised two 1D convolutions with a kernel size of 3 and group normalization with 8 groups, whereas the residual branch employed a 1D convolution with a kernel size of 1 for dimensionality projection.
In each Transformer block, the self-attention used two attention heads with a head dimension of 64, and the feed-forward layers used a dropout rate of 0.05. 
The time step $t$ was encoded as a 256-dimensional embedding via a sinusoidal positional embedding followed by a two-layer MLP, and the resulting time embedding was injected into all ResNet blocks and Transformer blocks.
For resolution changes, each downsampling layer was implemented using a 1D convolution with kernel size of 3, while each upsampling layer was implemented using a transposed 1D convolution with kernel size of 4; both used a stride of 2.
Finally, the output head consisted of two 1D convolutions with kernel sizes of 3 and 1, respectively, where the latter used the same number of output channels as the mel-spectrogram dimension to produce the velocity field.
}
The velocity field estimation network was trained on 1-second segments with batch size 48 for a total of 1.15M steps. 
The training process was divided into two phases. 
In the first 1.0M steps, we performed a straightforward training stage in which only the straightforward CFM loss is applied by setting $\lambda_{\text{self-cons}}=0$ and $\lambda_{\text{CFM}}=45$. 
In the subsequent 0.15M steps, we entered the self-consistency training stage, where self-consistency regularization was enabled with $\lambda_{\text{self-cons}}=10$ while $\lambda_{\text{CFM}}$ was kept unchanged.
In the self-consistency phase, we set $\epsilon=0.01$, $\Delta t_{\min}=0.005$, $\Delta t_{\max}=0.02$, and $\sigma=0.3$. 
This schedule stabilized velocity learning before
imposing time-invariance constraints for fast sampling. 
The training also adopted an AdamW optimizer with $(\beta_1,\beta_2)=(0.8,0.99)$ and an initial learning rate of $2\times10^{-4}$, which was exponentially decayed by a factor of $0.999$ per epoch. 
At inference time, the ODE is solved using only 4 iterations (i.e, $I=4$) to reduce computational cost.

\item \textbf{Configurations of Vocoding-driven Waveform Reconstruction Stage.}
We employed {HiFi-GAN\_v1~\cite{kong2020hifi}} as the neural vocoder in this stage and used its official configuration\footnote{\url{https://github.com/jik876/hifi-gan}.}, which was reimplemented and trained on our datasets {for 1M steps with a batch size of 16}.

\end{itemize}

\begin{table*}[!t]
\centering
\small
\renewcommand{\arraystretch}{1.02}
\setlength{\tabcolsep}{4.5pt}
\caption{Objective and subjective results of FMelCodec and baselines on the LibriTTS (16~kHz) and VCTK (48~kHz) test sets, evaluated at 250~bps and 750~bps, respectively. \textbf{Bold} and \underline{underline} numbers denote the best and second-best results, respectively.}
\resizebox{\linewidth}{!}{%
\label{tab1}
\begin{tabular}{l|ccccccc|ccccccc}
\hline

\hline
\multirow{2}{*}{Methods}
& \multicolumn{7}{c|}{LibriTTS (16 kHz, 250 bps)}
& \multicolumn{7}{c}{VCTK (48 kHz, 750 bps)} \\
\cline{2-15}
& ViSQOL$\uparrow$ & UTMOS$\uparrow$ & SIM$\uparrow$& dWER \textcolor{blue}{($\%$)}$\downarrow$& \textcolor{blue}{MCD (dB)}$\downarrow$ & NMOS$\uparrow$ & SMOS$\uparrow$
& ViSQOL$\uparrow$ & UTMOS$\uparrow$ & SIM$\uparrow$& dWER \textcolor{blue}{($\%$)}$\downarrow$ & \textcolor{blue}{MCD (dB)}$\downarrow$& NMOS$\uparrow$ & SMOS$\uparrow$\\
\hline
DAC \cite{kumar2024high}
& 2.79 & 1.96 & 0.86&{72.58} &	\textcolor{blue}{4.66}& {2.37$\pm$0.06} & {2.26$\pm$0.05}
& 3.28 & 2.98 & 0.88 &{33.07}&\textcolor{blue}{3.45}& 3.23$\pm$0.06 & 3.24$\pm$0.05
 \\
MDCTCodec \cite{jiang2024mdctcodec}
& \underline{3.45} & 2.43 & \textbf{0.92} &{29.27}&\textbf{\textcolor{blue}{3.32}}& {3.13$\pm$0.05} & {3.11$\pm$0.05}
& \underline{3.48} & 3.33 & \underline{0.92}&{\underline{9.62}} &\textcolor{blue}{3.13}& 3.48$\pm$0.05 & \underline{3.73$\pm$0.05}
 \\
BigCodec \cite{xin2024bigcodec}
& 3.22 & \underline{3.26} & \underline{0.90} &{41.26}& 	\textcolor{blue}{3.73}&{\textbf{3.74$\pm$0.05}} & {3.17$\pm$0.05}
& 3.34 & \textbf{3.67} & 0.91 &{10.84}&\textcolor{blue}{\underline{2.87}}& \textbf{3.75$\pm$0.05} & 3.72$\pm$0.05
 \\
WavTokenizer \cite{jiwavtokenizer}
& 2.61 & 1.94 & 0.84 & {73.97}&\textcolor{blue}{4.74}&{2.76$\pm$0.05} & {2.55$\pm$0.05}
& 3.32 & 2.34 & 0.75& {81.76} &\textcolor{blue}{4.41}& 3.52$\pm$0.05 & 3.30$\pm$0.05
 \\
FlowDec \cite{welkerflowdec}
& 2.38 & 1.32 & 0.84 &{76.36}&\textcolor{blue}{5.28}&{1.97$\pm$0.06} & {2.03$\pm$0.06}
& 2.77 & 2.93 & 0.87&{34.42} & \textcolor{blue}{3.82}&3.30$\pm$0.06 & 3.17$\pm$0.05
\\
{FocalCodec \cite{dellafocalcodec}}
& {3.12} & {\underline{3.26}} & {\textbf{0.92}} &\textcolor{blue}{\textbf{{4.97}}}&\textcolor{blue}{4.35}& {3.65$\pm$0.05} & {\underline{3.31$\pm$0.05}}
& {-} & {-} & {-} & {-} & {-}& {-}& {-}
\\
\hline
FMelCodec
& \textbf{3.56} & \textbf{3.48} & \textbf{0.92} & \textcolor{blue}{\underline{27.01}} &\underline{\textcolor{blue}{3.60}}& {\underline{3.72$\pm$0.05}}& {\textbf{3.51$\pm$0.05}}
& \textbf{3.62} & \underline{3.66} & \textbf{0.93} &{\textbf{4.80}}&\textbf{\textcolor{blue}{2.52}}&\underline{3.73$\pm$0.05} & \textbf{3.76$\pm$0.05}
 \\
\hline

\hline
\end{tabular}%
}
\end{table*}

\subsection{Baselines}
\label{subsec:baseline}

For a comprehensive comparison, we evaluated FMelCodec\footnote{Speech samples can be found at \url{https://redmist328.github.io/FMelCodec}. {Source code and checkpoint can be found at \url{https://github.com/redmist328/FMelCodec}.}} against several representative neural speech codecs that support operation at low bitrates. 
The details of the compared codecs are as follows.
\begin{itemize}[leftmargin=*]
\item \textbf{DAC}\footnote{\url{https://github.com/descriptinc/descript-audio-codec}.} \cite{kumar2024high}: 
DAC is a waveform-domain neural speech codec built upon a convolutional encoder--decoder and RVQ. 
It directly discretizes and reconstructs speech in the time domain, serving as a strong and widely used baseline.

\item \textbf{MDCTCodec} \cite{jiang2024mdctcodec}: 
MDCTCodec is a spectral-domain neural speech codec built upon a convolutional encoder--decoder and RVQ. 
It uses the MDCT spectrum as the target for discretization and reconstruction, and then recovers the decoded speech via inverse MDCT.

\item \textbf{BigCodec}\footnote{\url{https://github.com/Aria-K-Alethia/BigCodec}.} \cite{xin2024bigcodec}: 
BigCodec is a waveform-domain neural speech codec that employs large-capacity neural networks with single-VQ-based discretization to achieve high-fidelity reconstruction, at the cost of increased model size and computational complexity compared to deployment-oriented designs.

\item \textbf{WavTokenizer}\footnote{\url{https://github.com/jishengpeng/WavTokenizer}.} \cite{jiwavtokenizer}: 
WavTokenizer is an asymmetric neural speech codec that encodes speech in the waveform domain and decodes it in the STFT domain. 
It adopts a single VQ with a large codebook for discretization and employs a more powerful decoder architecture to achieve high-quality speech coding at low bitrates.

\item \textbf{FlowDec}\footnote{\url{https://github.com/facebookresearch/FlowDec}.}~\cite{welkerflowdec}: 
FlowDec combines an RVQ-based DAC-style waveform-domain neural speech codec with a CFM-based post-processing module, which further enhances the STFT spectrum of the decoded speech to improve overall coding quality.

\item {\textbf{FocalCodec}\footnote{\url{https://github.com/lucadellalib/focalcodec}.} \cite{dellafocalcodec}: 
FocalCodec is an ultra-low-bitrate neural speech codec that leverages an SSL-based speech encoder, i.e., WavLM~\cite{chen2022wavlm}, together with a compressor--quantizer--decompressor architecture based on focal modulation. 
It employs a single binary codebook for discretization and supports ultra-low-bitrate speech coding.}

\item  {\textbf{SemantiCodec}\footnote{\url{https://github.com/haoheliu/SemantiCodec-inference}.} \cite{liu2024semanticodec}: 
SemantiCodec is an ultra-low-bitrate neural speech codec that combines a pre-trained SSL-based semantic encoder and an acoustic encoder to capture both high-level semantics and residual acoustic details, while employing a diffusion-based decoder for reconstruction.}

\end{itemize}


{For a more comprehensive evaluation, we reported two types of baseline comparisons.} 
{We first compared FMelCodec with several representative baselines including DAC, MDCTCodec, BigCodec, WavTokenizer, FlowDec and FocalCodec, under the same ultra-low-bitrate budget. 
In this setting, all baselines were retrained based on their official open-source implementations, with their quantization and temporal downsampling configurations adapted when necessary to match our bitrate requirement. 
While keeping their original quantization type unchanged, they all adopted a single codebook containing 1024 entries. 
Regarding the temporal downsampling configuration, for DAC, BigCodec, WavTokenizer, and FlowDec, we set the overall temporal downsampling factor to 640 by adopting downsampling ratios of [8, 5, 4, 4].
For MDCTCodec, we set the MDCT hop size to 160 and the model downsampling factor to 4, resulting in an overall temporal downsampling factor of 640.
For FocalCodec, we followed the temporal downsampling configuration used in its 330-bps version, except that the codebook size was reduced to 1024.
Under the current configuration, the baselines operated at 250 bps for 16 kHz and 750 bps for 48 kHz. 
FocalCodec was excluded from the 48-kHz experiment because it relies on a pretrained SSL encoder designed for 16 kHz speech input and thus does not directly support 48 kHz speech coding.
Since the training code of SemantiCodec is not publicly available, it was not included in this experiment.
}

{We additionally compared FMelCodec with baselines for which ultra-low-bitrate public checkpoints are available. 
Both FocalCodec and SemantiCodec provided checkpoints for 16 kHz speech (no 48-kHz checkpoints). 
At 16 kHz, we compared the proposed FMelCodec at 250 bps with FocalCodec at 330 bps (denoted as FocalCodec$^\dagger$) and SemantiCodec at 310 bps (denoted as SemantiCodec$^\dagger$). 
DAC, BigCodec, and FlowDec were excluded from this experiment because their available checkpoints correspond to bitrates far above the ultra-low-bitrate regime targeted by FMelCodec. MDCTCodec and WavTokenizer were also excluded, as no 16 kHz or 48 kHz checkpoints are publicly available for them.
}


\subsection{Evaluation Metrics}
\label{subsec:eval_metrics}

We adopt both objective and subjective metrics to evaluate FMelCodec and the baseline methods. 

\begin{itemize}[leftmargin=*]
\item 
\textbf{Objective Metrics.}
We reported three objective perceptual metrics for reconstructed speech quality, i.e., ViSQOL~\cite{chinen2020visqol}, UTMOS~\cite{saeki2022utmos}, and speaker similarity (SIM). 
ViSQOL is an intrusive metric that evaluates speech fidelity in a perceptually motivated feature space, whereas UTMOS is a non-intrusive neural mean opinion score (MOS) predictor that correlates with perceived speech naturalness.
SIM is an intrusive metric that measures speaker similarity by comparing speaker embeddings extracted from the reference and reconstructed utterances. 
Specifically, we use a WavLM-based speaker verification model (WavLM-TDNN4)~\cite{chen2022wavlm} initialized from the \texttt{wavlm\_large\_finetune}  checkpoint to extract embeddings, and compute the cosine similarity between them.
{We further reported differential word error rate (dWER) as a content-preservation metric.
Specifically, we used a Whisper-based \cite{radford2023robust} automatic speech recognition (ASR) model to transcribe both the reconstructed and reference utterances, and then computed the WER between the two recognized texts.}
\textcolor{blue}{We also reported mel cepstral distortion (MCD) as a frequency-domain distortion metric which measures the distance between the mel-cepstral representations of the reference and reconstructed speech.}
\textcolor{blue}{In terms of efficiency and complexity, we reported three objective metrics, i.e., the real-time factor (RTF), the floating-point operations (GFLOPs) required to reconstruct 1 second of speech, and the number of model parameters (Param.). 
The RTF, defined as the ratio between the speech generation time and the duration of the generated speech, was computed by averaging over the entire test set on a single NVIDIA A100 GPU, and is used to assess practical inference speed. 
GFLOPs and Param. were used to assess computational complexity and model complexity, respectively.}


\item 
\textbf{Subjective Metrics.}
When comparing FMelCodec with other baselines at equal {bitrates}, due to the involvement of multiple codecs, we conducted naturalness MOS (NMOS) and speaker-similarity MOS (SMOS) tests on Amazon Mechanical Turk (AMT)\footnote{\url{https://www.mturk.com/}.}. 
For each MOS evaluation, we constructed a set of 20 utterances and collected ratings from at least 30 listeners.
In the NMOS test, participants rated naturalness on a 5-point scale with 1-point increments. 
In the SMOS test, participants rated how similar the reconstructed speaker sounds compared to a natural reference, using the same 5-point scale. 
We also computed the 95\% confidence intervals of the MOS scores for statistical analysis.
When comparing FMelCodec with \textcolor{blue}{equal-bitrate strong baselines}, higher-bitrate baselines and performing analysis experiments on FMelCodec, we conducted ABX preference tests to enable more targeted comparisons between two systems.
For each ABX test, we prepared 20 pairs of A/B utterances and recruited at least 30 AMT listeners to select the preferred sample or indicate ``no preference''.
We also report the $p$-values of paired $t$-tests for statistical analysis.

\end{itemize}

\begin{figure*}
    \centering
    \includegraphics[width=1\linewidth]{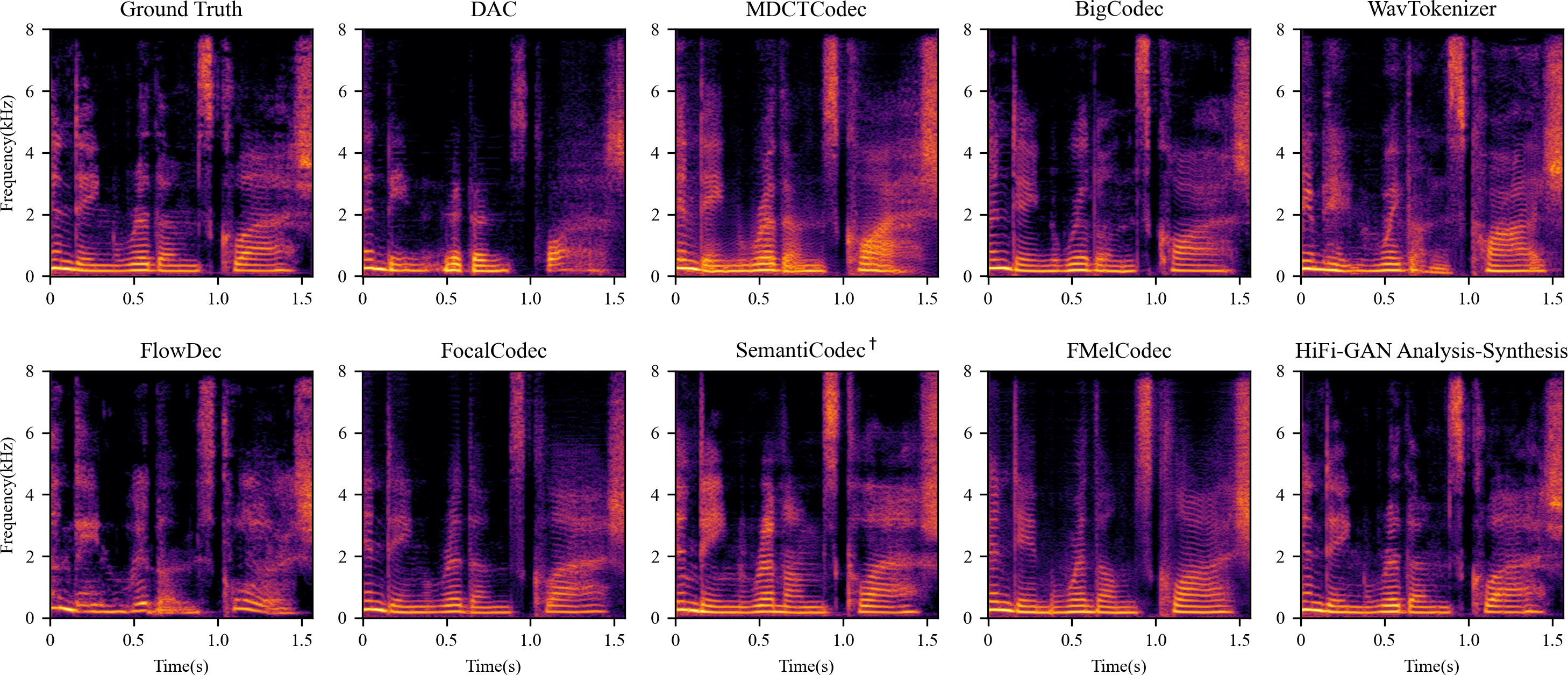}
\caption{{Spectrogram visualizations of natural speech, speech decoded by FMelCodec and the baseline methods, as well as speech reconstructed via HiFi-GAN analysis–synthesis. The example corresponds to the same utterance from the 16~kHz LibriTTS test set, evaluated at 250~bps except for SemantiCodec$^\dagger$ which is evaluated at 310 bps.}}
    \label{fig:mainyupu}
\end{figure*}

\vspace{-2mm}

\subsection{Comparison at Equal Ultra-Low Bitrates}
\label{subsec:Comparison at Equal Ultra-Low Bitrates}

To verify that the proposed FMelCodec can maintain high coding performance under ultra-low bitrate constraints, we compared it with {representative} baselines at matched ultra-low bitrates, i.e., 250~bps for 16~kHz LibriTTS and 750~bps for 48~kHz VCTK. 
\textcolor{blue}{Both quality-related objective metrics and subjective listening-test results are reported in Table~\ref{tab1}. 
Efficiency- and complexity-related metrics are reported in Table~\ref{tabadd}.}
For simplicity, \textcolor{blue}{efficiency- and} complexity-related metrics were measured only on the LibriTTS dataset, since absolute values may vary across datasets (sampling-rate dependent) but the relative trends and system-wise comparisons remain consistent.
{For FMelCodec, the reported \textcolor{blue}{RTF,} GFLOPs and Param. in \textcolor{blue}{Table~\ref{tabadd}} accounted for all three stages of the CRR framework.
}
In addition, to facilitate qualitative analysis, spectrogram visualizations of the decoded speech from different codecs are provided in Fig.~\ref{fig:mainyupu}.

We first examine the 16~kHz, 250~bps setting, which represents the most challenging ultra-low-bitrate regime. As listed in Table~\ref{tab1}, DAC failed at this bitrate, with both objective metrics and subjective MOS scores dropping to unacceptable levels, highlighting the difficulty of preserving fine-grained temporal structure under extreme waveform-domain compression. 
MDCTCodec, as an MDCT-spectral baseline, achieved a clear performance improvement over the waveform-domain DAC, but still fell short of FMelCodec across multiple objective and subjective metrics, despite its advantage in \textcolor{blue}{efficiency and complexity}. 
BigCodec achieved competitive NMOS only at the expense of substantially higher model complexity, using nearly six times as many parameters as FMelCodec, while its speaker similarity still remained inferior to FMelCodec according to the SMOS results. 
{Overall, considering both reconstruction quality, \textcolor{blue}{efficiency} and complexity, FMelCodec provides a more favorable trade-off than BigCodec.}
WavTokenizer exhibited a noticeable performance degradation at 250~bps, while FlowDec performed worst due to its DAC-style bottleneck and prohibitive computational cost. 
{FocalCodec, which is built upon pretrained SSL speech representations, exhibited a different behavior from the above baselines. 
At 250~bps, it achieved a markedly lower dWER than all other methods (including the proposed FMelCodec), which is consistent with its representation design, since SSL features are more effective at preserving speech content information; however, it was clearly inferior to FMelCodec on the other quality-related metrics.
This also reflects a key difference from the other methods, which are designed as purely acoustic-level codecs, whereas FocalCodec additionally benefits from semantic information carried by pretrained SSL representations.
However, such semantic information comes at the cost of higher model complexity, as it requires large-scale pretrained SSL models. 
In contrast, the proposed FMelCodec demonstrated stable reconstruction performance at only 250~bps, achieving MOS scores above 3.5 in both naturalness and speaker similarity, which indicates that the decoded speech is not only of high quality but also preserves speaker identity effectively. 
\textcolor{blue}{We further conducted a subjective ABX preference test between FMelCodec and FocalCodec. 
The preference rates for FMelCodec, no preference, and FocalCodec were 36.21\%, 31.52\%, and 32.27\%, respectively, and the difference was not statistically significant ($p=0.22$). 
This result further indicates that FMelCodec achieved perceived naturalness comparable to FocalCodec, while preserving a purely acoustic-level design with substantially higher efficiency and lower complexity, consistent with the NMOS results.}
Moreover, FMelCodec achieved a substantially lower dWER than all purely acoustic baselines, indicating that, even without the aid of semantic representations, it is still able to preserve speech content information to some extent.}
These observations are further corroborated by the spectrogram visualizations in Fig.~\ref{fig:mainyupu}, which show that FMelCodec preserved harmonic structure and pitch stability more faithfully than the baseline methods. Several baselines (e.g., DAC, MDCTCodec, and FlowDec) clearly struggled to reconstruct fine spectral details and exhibited disrupted pitch trajectories, which may in turn lead to pronunciation errors. 
{Moreover, FMelCodec maintained acceptable computational and model complexity compared to the baseline methods, especially when contrasted with high-capacity methods such as BigCodec and FocalCodec.}
\begin{table}[!t]
\centering
\small
\renewcommand{\arraystretch}{1.02}
\caption{\textcolor{blue}{Efficiency and complexity comparison of FMelCodec and baseline codecs on the LibriTTS (16 kHz) test set. Here, ``$a\times$" represents $a\times$ real time.}}
\resizebox{0.89\linewidth}{!}{%
\label{tabadd}
\begin{tabular}{l|ccc}
\hline

\hline
\textcolor{blue}{Methods}
& \textcolor{blue}{RTF$\downarrow$}
& \textcolor{blue}{GFLOPs$\downarrow$}
& \textcolor{blue}{Param. (M)$\downarrow$}\\
\hline
\textcolor{blue}{DAC \cite{kumar2024high}}
& \textcolor{blue}{0.096 (10.41$\times$)}
& \textcolor{blue}{32.22}
& \textcolor{blue}{73.96}
\\
\textcolor{blue}{MDCTCodec \cite{jiang2024mdctcodec}}
& \textcolor{blue}{\textbf{0.013 (75.29$\times$)}}
& \textcolor{blue}{\textbf{2.49}}
& \textcolor{blue}{\textbf{6.61}}
\\
\textcolor{blue}{BigCodec \cite{xin2024bigcodec}}
& \textcolor{blue}{0.052 (19.07$\times$)}
& \textcolor{blue}{28.03}
& \textcolor{blue}{158.31}
\\
\textcolor{blue}{WavTokenizer \cite{jiwavtokenizer}}
& \textcolor{blue}{\underline{0.021 (47.66$\times$)}}
& \textcolor{blue}{\underline{4.21}}
& \textcolor{blue}{71.65}
\\
\textcolor{blue}{FlowDec \cite{welkerflowdec}}
& \textcolor{blue}{0.214 (4.67$\times$)}
& \textcolor{blue}{2280}
& \textcolor{blue}{97.64}
\\
\textcolor{blue}{FocalCodec \cite{dellafocalcodec}}
& \textcolor{blue}{0.026 (38.13$\times$)}
& \textcolor{blue}{8.84}
& \textcolor{blue}{143.30}
\\
\hline
\textcolor{blue}{FMelCodec}
& \textcolor{blue}{0.022 (44.82$\times$)}
& \textcolor{blue}{18.47}
& \textcolor{blue}{\underline{27.17}}
\\
\hline

\hline
\end{tabular}%
}
\end{table}

\begin{table*}[!t]
\centering
\small
\renewcommand{\arraystretch}{1.2}
\caption{
Objective and subjective results of FMelCodec at 250 bps, FocalCodec$^\dagger$ at 330 bps and SemantiCodec$^\dagger$ at 310 bps on the LibriTTS (16 kHz) test set.
}
\resizebox{0.98\linewidth}{!}{%
\label{tabadd2}
\begin{tabular}{l|ccccccccccc}
\hline

\hline
{{Methods}}
& {Bitrate (bps)$\downarrow$} &{ViSQOL$\uparrow$} & {UTMOS$\uparrow$} & {SIM$\uparrow$}& {dWER \textcolor{blue}{($\%$)}$\downarrow$}&\textcolor{blue}{{MCD (dB)$\downarrow$}}& {NMOS$\uparrow$} & {SMOS$\uparrow$ }&{{RTF$\downarrow$}}&{{GFLOPs$\downarrow$}}&{{Param. (M)$\downarrow$}}\\
\hline

{FocalCodec$^\dagger$  \cite{dellafocalcodec}}
& {330} &{3.49} & {4.09} & {{0.95}} &{3.21}& \textcolor{blue}{3.70}&{3.86$\pm$0.06}& {3.78$\pm$0.06}&0.014 (69.43$\times$)&{8.84} & {143.30} 
\\
{SemantiCodec$^\dagger$ \cite{liu2024semanticodec}}& {310}
& {3.32} & {2.62} & {{0.91}} &{44.82}& \textcolor{blue}{4.24}& {3.21$\pm$0.07}& {3.52$\pm$0.06}&3.267 (0.30$\times$)&{1599} & {1033} 
\\
\hline
{FMelCodec}& {250}
& {{3.56}} & {{3.48}} & {{0.92}} & {27.01} &\textcolor{blue}{3.60}& {3.79$\pm$0.06}& {3.88$\pm$0.06}&0.022 (44.82$\times$)& {18.47} & {{27.17}}
 \\
\hline

\hline
\end{tabular}}
\end{table*}

At 48~kHz and 750~bps, all codecs exhibited improved objective and subjective performance, with MOS scores generally falling within an acceptable range. This improvement can be attributed to the higher absolute bitrate, which moderately relaxes the reconstruction difficulty despite the matched compression ratio across sampling rates. Nevertheless, the relative behaviors observed at 16~kHz largely persist: the proposed FMelCodec continued to significantly outperform DAC, MDCTCodec, WavTokenizer, and FlowDec in terms of reconstructed speech naturalness, speaker similarity, \textcolor{blue}{content preservation, and frequency-domain distortion}, while achieving performance comparable to or better than the much more complex BigCodec. 
{FocalCodec does not support 48-kHz speech coding, which further highlights the limited flexibility of such SSL-based codecs for higher-sampling-rate speech coding.
}

{Based on the above analysis, we summarize the advantages and limitations of FMelCodec as follows. 
FMelCodec is a purely acoustic-level speech codec that exhibits strong overall speech reconstruction performance in terms of naturalness and similarity under extremely low-bitrate constraints, while maintaining \textcolor{blue}{real-time inference efficiency and} acceptably low computational and model complexity, thereby achieving a favorable balance between reconstruction quality and complexity. 
However, to keep the complexity under control, FMelCodec does not incorporate semantic features as auxiliary information. 
Consequently, it still lags behind SSL-based codecs in terms of content preservation, which we consider an important direction for future improvement.
}

\vspace{-2mm}
\subsection{{Comparison with Baselines using Publicly Available Checkpoints}}
\label{subsubsec:ee_bitrate_comparison2}

{To complement the matched-bitrate retraining results, we further compared FMelCodec at 250 bps with FocalCodec$^\dagger$ at 330 bps and SemantiCodec$^\dagger$ at 310 bps for 16-kHz speech coding.
The objective and subjective experimental results are summarized in Table~\ref{tabadd2}. 
}

{
We can see that FocalCodec$^\dagger$, operating at its native 0.33~kbps, surpassed FMelCodec on several objective metrics, including UTMOS, SIM, and dWER. 
However, these objective gains did not translate into a clearly perceptible advantage in subjective evaluation.
According to the NMOS and SMOS results, FMelCodec and FocalCodec$^\dagger$ exhibited comparable perceptual performance. 
For example, the difference in NMOS was not statistically significant according to a paired $t$-test ($p=0.13$). 
Regarding the SMOS, FMelCodec even slightly outperformed FocalCodec$^\dagger$ ($p<0.05$), suggesting better preservation of speaker timbre. 
Although FocalCodec achieved a slightly better SIM score, this may be attributed to the fact that both its input and the SIM calculation rely on features extracted by WavLM.
The above results indicated that, from the perspective of speech reconstruction quality, FMelCodec remained competitive with FocalCodec$^\dagger$ despite being a purely acoustic-level codec. 
Moreover, FMelCodec operated at a lower bitrate and used a much smaller model size, suggesting a more favorable quality–complexity trade-off.
It is also worth noting that the official FocalCodec$^\dagger$ was trained on the full LibriTTS corpus (approximately 586 hours), whereas FMelCodec was trained only on the train-clean-100 and train-clean-360 subsets (approximately 245 hours). 
The performance gap between FocalCodec$^\dagger$ in Table~\ref{tabadd2} and the retrained FocalCodec in Table~\ref{tab1} further suggests that increasing bitrate together with larger data scale is beneficial to codec performance.
}

{
By contrast, FMelCodec outperformed SemantiCodec$^\dagger$ on all reported objective and subjective metrics, which further supports the effectiveness of the proposed method. 
This observation is also consistent with the spectrogram visualization in Fig.~\ref{fig:mainyupu}, where the speech spectrogram generated by SemantiCodec$^\dagger$ shows noticeably weaker harmonic reconstruction (e.g., in the 1.0--1.5~s region). 
In addition, SemantiCodec employed a diffusion-based decoder, which resulted in extremely high computational complexity and inference cost, thereby making practical deployment much more difficult.
Overall, these comparisons demonstrated that FMelCodec achieved a strong balance among perceptual quality, bitrate, and complexity, even when compared with competitive baselines for which public checkpoints are available.
}
\vspace{-2mm}
\begin{figure}[t]
    \centering
    \includegraphics[width=1\linewidth]{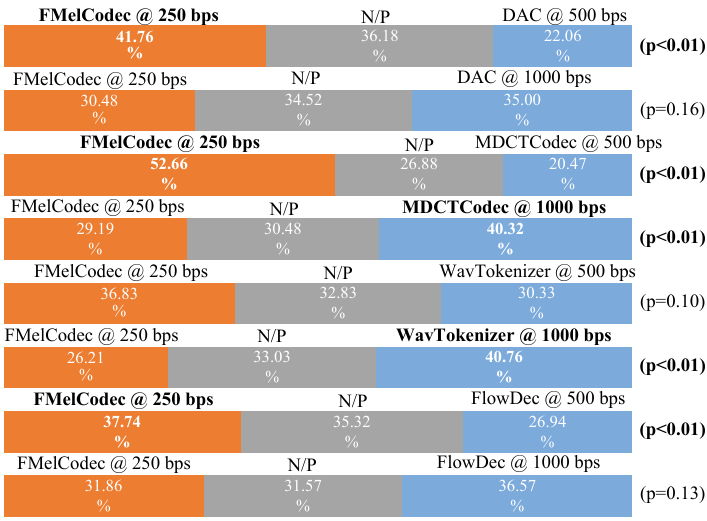}
    \caption{ABX preference results on the 16 kHz LibriTTS test set, comparing FMelCodec at 250~bps with baseline codecs operated at 500/1000~bps. ``N/P'' denotes no preference and $p$ indicates the $p$-value of a paired $t$-test used for statistical analysis.}
    \label{fig:abx_500bps}
\end{figure}
\vspace{-1mm}
\subsection{Comparison against Higher-Bitrate Baselines}
\label{subsubsec:high_bitrate_comparison2}

Building on the results in the previous section, which demonstrate that FMelCodec outperformed the baselines under equal ultra-low-bitrate constraints, we further conducted a quantitative analysis to measure the degree of this advantage in terms of bitrate savings. 
To explicitly quantify the bitrate savings enabled by FMelCodec, we compared FMelCodec operating at 250~bps for 16~kHz speech against several baseline codecs running at substantially higher bitrates. 
Concretely, we evaluated DAC, MDCTCodec, WavTokenizer, and FlowDec at 500~bps and 1000~bps, and directly compared them with FMelCodec at 250~bps. 
{BigCodec and FocalCodec are} excluded from this analysis, since results in Section~\ref{subsec:Comparison at Equal Ultra-Low Bitrates} indicate that under the same 250~bps constraint, {BigCodec and FocalCodec} attained perceptual quality comparable to FMelCodec thanks to the larger model footprint. 
To enable focused subjective comparisons between pairs of systems, we conducted ABX preference tests, in which FMelCodec at 250~bps is evaluated against each higher-bitrate baseline individually. 
The corresponding results are summarized in Fig.~\ref{fig:abx_500bps}.

The ABX results revealed clear bitrate-saving effects. 
FMelCodec at 250~bps was consistently preferred ($p<0.01$) over DAC and FlowDec even when these waveform-domain codecs operate at 500~bps, and became statistically indistinguishable ($p>0.01$) from them when their bitrates are further increased to 1000~bps. 
These results indicate that FMelCodec can achieve a similar level of perceived quality while saving at least 750~bps compared with these RVQ-based waveform codecs. 
Comparison against MDCTCodec exhibited a different behavior: compared with FMelCodec at 250~bps, MDCTCodec at 500~bps remained less preferred ($p<0.01$), while at 1000~bps it showed a clear ABX advantage ($p<0.01$). 
This indicated that MDCT-spectral-based codecs were strongly bitrate-sensitive. 
Based on these results, FMelCodec was shown to save at least 250~bps, but less than 750~bps, relative to MDCTCodec. 
WavTokenizer represented a stronger low-bitrate baseline. 
When compared against FMelCodec at 250~bps, ABX results were inconclusive ($p>0.01$) when WavTokenizer operates at 500~bps, whereas listeners tended to prefer WavTokenizer at 1000~bps ($p<0.01$). 
This indicates that FMelCodec can save at most 250~bps relative to WavTokenizer. 
Overall, these findings demonstrate that FMelCodec provided an efficient solution for ultra-low-bitrate speech coding, achieving competitive perceptual quality while saving approximately 250$\sim$750~bps compared with baseline neural codecs.

\begin{figure}[t]
    \centering
    \includegraphics[width=1\linewidth]{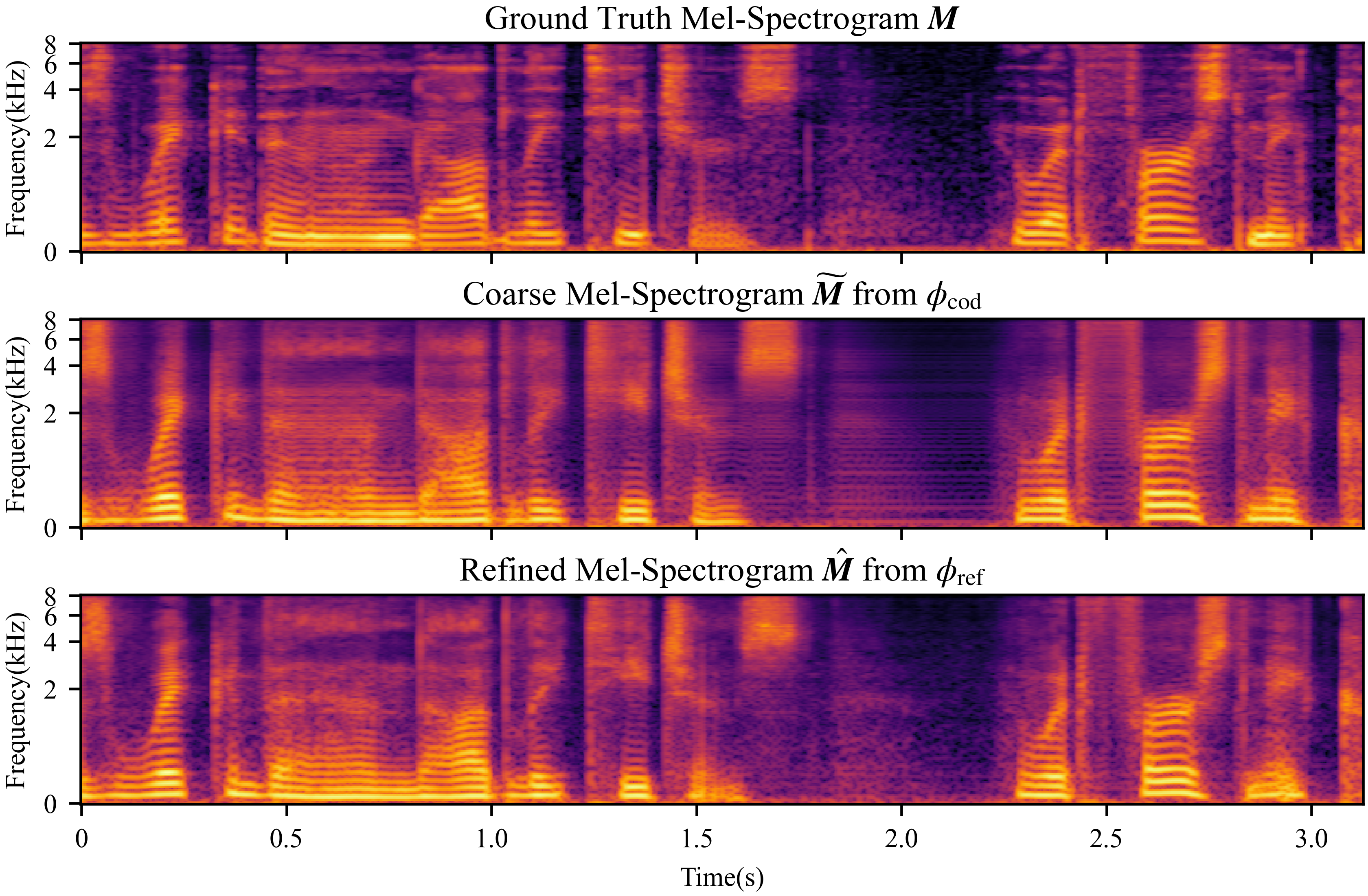}
    \caption{Visualization of the natural mel-spectrogram $\bm{M}$ and the mel-spectrograms $\widetilde{\bm{M}}$ and $\hat{\bm{M}}$ produced by stage $\phi_{\text{cod}}$ and $\phi_{\text{ref}}$, respectively.}
    \label{fig:bwe3}
\end{figure}


\begin{table}[t!]
\centering
\caption{Complexity analysis results of the three individual stages in FMelCodec. The numbers in $(\cdot)$ indicate the proportion contributed by each stage to the overall system.}
\label{tab:high_bitrate_obj}
\begin{tabular}{l|ccc}
\hline

\hline
 & $\phi_{\text{cod}}$ & $\phi_{\text{ref}}$ & $\phi_{\text{voc}}$ \\
\hline
GFLOPs & 0.60 (3.25\%) & 1.48 (8.02\%) & 16.38 (88.73\%) \\
Param. (M) & 6.29 (23.15\%) & 7.84 (28.86\%) & 13.04 (47.99\%) \\
\hline

\hline
\end{tabular}
\end{table}

\subsection{Analysis and Discussion}

In this section, we move beyond baseline comparisons and provide an in-depth analysis and discussion of FMelCodec. 
The focus is placed on understanding the internal mechanisms of the proposed three-stage framework, clarifying the role of each stage, and identifying their respective strengths and limitations. 
We further investigate the effectiveness of several key design choices, including online clustering for VQ, stage-wise training, and self-consistency regularization for CFM. 
All experiments were conducted on the LibriTTS dataset (16~kHz) at a bitrate of 250~bps.




\subsubsection{\textbf{Role Analysis of Three Stages}}
The proposed FMelCodec adopts the CRR framework which consists of three cascaded stages, i.e., $\phi_{\text{cod}}$, $\phi_{\text{ref}}$, and $\phi_{\text{voc}}$. 
A shared characteristic of these three stages is that they all operate on mel-spectrogram representations, forming a unified processing pipeline centered on the mel-spectral domain. 
Therefore, to facilitate a qualitative analysis of the mel-spectrogram evolution across stages, we visualized the natural mel-spectrogram $\bm{M}$ together with the mel-spectrograms $\widetilde{\bm{M}}$ and $\hat{\bm{M}}$ produced by $\phi_{\text{cod}}$ and $\phi_{\text{ref}}$, respectively, as shown in Fig.~\ref{fig:bwe3}. 
Compared with the natural mel-spectrogram, the output of stage $\phi_{\text{cod}}$ is visibly coarse, but it still preserves the essential spectral structure of the speech signal. 
The stage $\phi_{\text{ref}}$ then effectively restores and sharpens spectral details, improving overall mel-spectrogram quality and bringing it closer to the natural target. 
From the spectrograms of FMelCodec and the HiFi-GAN analysis–synthesis results (i.e., driven by natural mel-spectrograms) in Fig.~\ref{fig:mainyupu}, it can be observed that the stage $\phi_{\text{voc}}$ is able to achieve high-quality and robust speech reconstruction even when driven by non-natural mel-spectrogram inputs.
These results confirm that the three stages fulfill their respective design objectives.

We further analyze the complexity of each stage, with the results summarized in Table~\ref{tab:high_bitrate_obj}. 
We can observe that FMelCodec’s advantage does not come from increasing the complexity of its core components. 
The two key stages of the CRR framework, i.e., $\phi_{\text{cod}}$ and $\phi_{\text{ref}}$ are lightweight in both model and computational complexity, with their inference FLOPs accounting for less than 12\% of the overall cost.
In contrast, stage $\phi_{\text{voc}}$ dominates the computational and parameter complexity, but it is not the focus of this work. 
This analysis further suggests that, within the CRR framework, exploring more lightweight neural vocoders constitutes a promising direction for future research.

\begin{figure}
    \centering
    \includegraphics[width=1\linewidth]{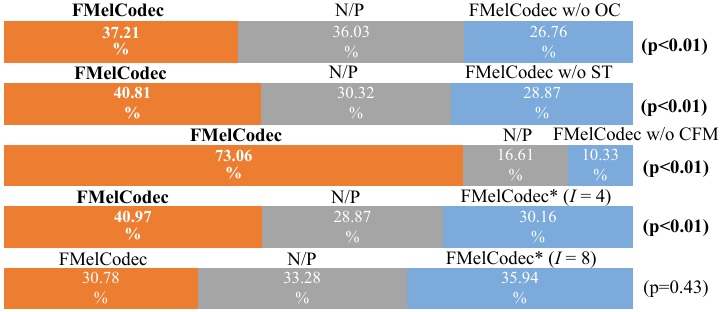}
\caption{\textcolor{blue}{ABX preference results on the 16 kHz LibriTTS test set, comparing FMelCodec at 250~bps with its ablated variants. ``N/P'' denotes no preference and $p$ indicates the $p$-value of a paired $t$-test used for statistical analysis.}}
\label{fig:ablation_abx}
\end{figure}


\subsubsection{\textbf{\textcolor{blue}{Ablation Studies}}}

To verify the effectiveness of the online clustering strategy used for quantization in stage $\phi_{\text{cod}}$, we ablated this component from FMelCodec by replacing the proposed OC-VQ with a plain VQ (denoted as FMelCodec w/o OC). 
We conducted ABX preference listening tests, and the results are shown in Fig.~\ref{fig:ablation_abx}. 
The ABX results in Fig.~\ref{fig:ablation_abx} show a statistically significant listener preference for the full FMelCodec over the w/o OC variant ($p<0.01$), confirming the importance of online clustering. 
Removing online clustering leads to severe codebook collapse in the VQ module, with only 359 out of 1024 codewords being effectively utilized (35.06\% utilization), whereas employing online clustering achieves full codebook utilization. 
This reduction in quantization capacity directly degrades the decoded coarse mel-spectrogram and substantially increases the difficulty of subsequent CFM-based refinement.



In the CRR framework of FMelCodec, the training of stages $\phi_{\text{cod}}$ and $\phi_{\text{ref}}$ is performed in a stage-wise and sequential manner, as illustrated in Fig.~\ref{fig:placeholder}. 
We ablated the stage-wise training strategy (denoted as FMelCodec w/o ST), i.e., jointly training stages $\phi_{\text{cod}}$ and $\phi_{\text{ref}}$. 
As shown by the ABX results in Fig.~\ref{fig:ablation_abx}, this setting consistently underperformed the proposed stage-wise pipeline and exhibited a significant listener preference gap in favor of FMelCodec ($p<0.01$). 
This behavior can be attributed to the moving-target issue introduced by one-stage joint training: as the codec parameters evolve, the resulting degradation patterns and paired data distribution become non-stationary, making it more difficult for the refinement stage to learn a stable and accurate velocity field.

\textcolor{blue}{We further ablated the CFM-based mel-spectrogram refinement stage $\phi_{\text{ref}}$ (denoted as FMelCodec w/o CFM) to examine the necessity of the refinement step in the CRR framework. 
In this variant, the decoded coarse mel-spectrogram $\widetilde{\bm{M}}$ produced by $\phi_{\text{cod}}$ was directly fed into the vocoder without CFM-based refinement. 
As shown in Fig.~\ref{fig:ablation_abx}, removing the refinement stage led to the largest preference gap among all ablated variants, with listeners strongly preferring the full FMelCodec over FMelCodec w/o CFM ($p<0.01$). 
This result indicates that, although the coding stage preserves main spectral structure under ultra-low-bitrate constraint, the decoded coarse mel-spectrogram is still insufficient for high-quality waveform reconstruction. 
The CFM-based refinement stage plays an essential role in recovering fine-grained mel-spectral details and reducing quantization-induced distortions before vocoding.}

\begin{figure}
    \centering
    \includegraphics[width=1\linewidth]{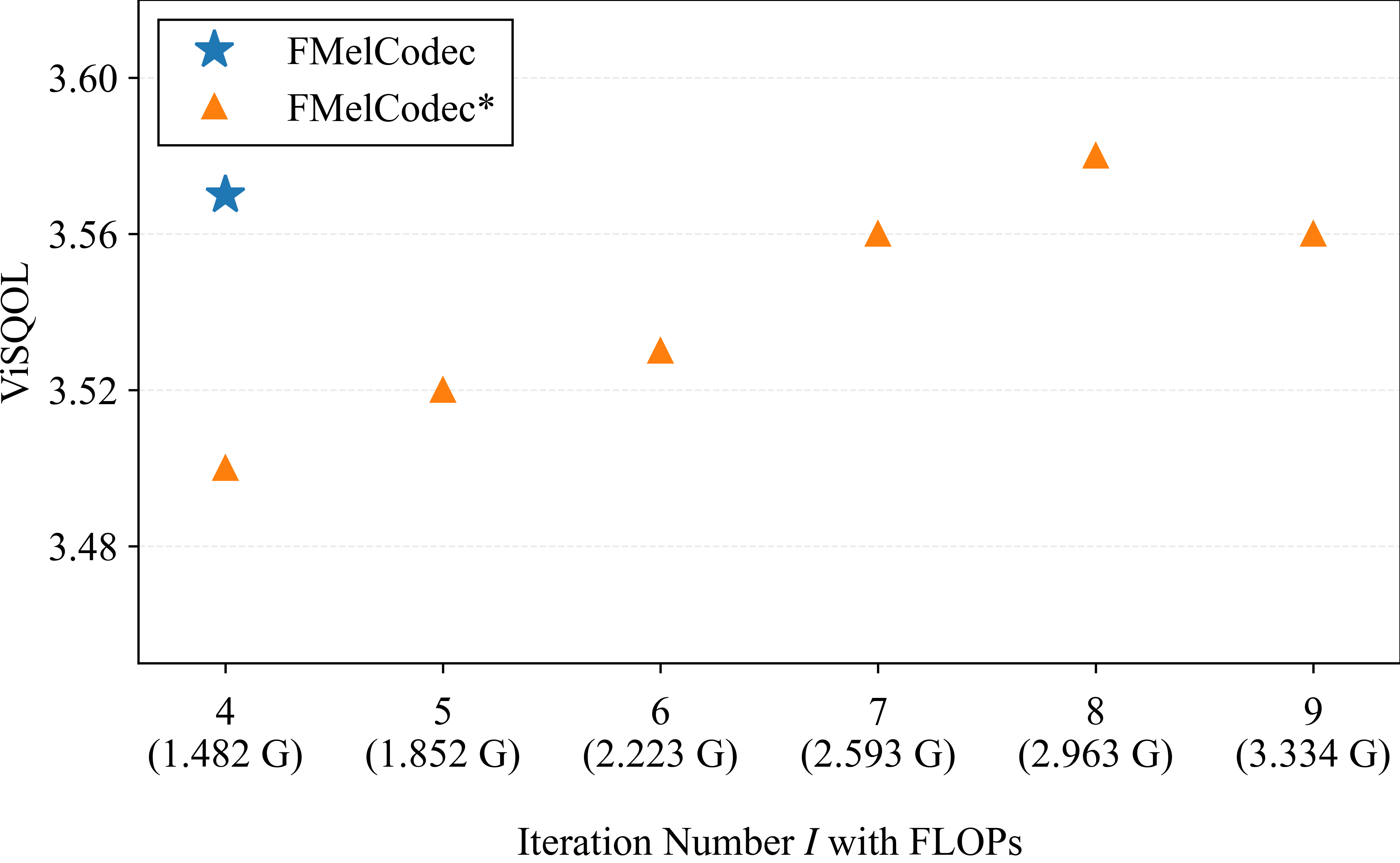}
\caption{Scatter plot of ViSQOL scores versus the number of CFM ODE iterations $I$ for FMelCodec and FMelCodec* at 250~bps on 16~kHz speech. The contents in $(\cdot)$ indicate the FLOPs for stage $\phi_{\text{ref}}$.}

    \label{fig:123}
\end{figure}

\subsubsection{\textbf{Effectiveness Analysis of Self-Consistency Loss}}

In the CFM-based mel-spectrogram refinement stage $\phi_{\text{ref}}$ of the CRR framework in FMelCodec, we introduced the self-consistency loss $\mathcal{L}_{\text{self-cons}}$ with the goal of reducing the number of ODE iterations $I$ required at inference time, thereby potentially lowering the computational cost. 
In the FMelCodec setting, we set $I=4$. 
To verify the effectiveness of the self-consistency loss, we constructed FMelCodec*, which removed this loss term, and evaluated it with $I$ set to $4,5,6,7,8,$ and $9$, respectively. 
We computed the ViSQOL scores of the generated speech and FLOPs of stage $\phi_{\text{ref}}$ for the above codecs and plotted them as the scatter plot shown in Fig.~\ref{fig:123}. 
For FMelCodec*, when the iteration number was the same ($i.e., I=4$), the speech quality was clearly inferior to that of FMelCodec with self-consistency according to the ViSQOL results. 
As the iteration number $I$ increased, the performance of FMelCodec* gradually improved, but the gain became marginal at larger $I$. 
Our proposed FMelCodec achieved a reconstruction quality at $I=4$ that was comparable to FMelCodec* using $I=8$ iterations, while requiring only about half of the computational cost according to the FLOPs.
\textcolor{blue}{To further examine whether this objective trend is perceptually meaningful, we also conducted ABX preference tests between FMelCodec and FMelCodec* under representative iteration settings. 
As shown in Fig.~\ref{fig:ablation_abx}, when both models used $I=4$, listeners significantly preferred FMelCodec over FMelCodec* ($p<0.01$), indicating that self-consistency brings a clear subjective improvement under the same low-step inference budget. 
When FMelCodec* increased the number of iterations to $I=8$, the ABX preference between FMelCodec and FMelCodec* became statistically insignificant ($p=0.43$), which is consistent with the ViSQOL--FLOPs analysis in Fig.~\ref{fig:123}. }
These results confirmed that the proposed self-consistency loss effectively reduced inference ODE iterations and computational cost.

\vspace{-2mm}
\section{Conclusion}
\label{sec:conclusion}
We presented FMelCodec, an ultra-low-bitrate neural speech codec operating in the mel-spectrogram domain under a three-stage CRR framework. 
In the mel-spectrogram coding stage, we employed a ConvNeXt v2 encoder–decoder with a single 1024-entry OC-VQ codebook and an effective waveform-level compression factor of 640$\times$, enabling operation at 250~bps for 16~kHz speech and 750~bps for 48~kHz speech. 
The subsequent CFM-based mel-spectrogram refinement stage leveraged a lightweight velocity-field estimator to enhance the decoded coarse mel-spectrogram, while the proposed self-consistency training scheme supported accurate inference with fewer ODE solver steps and reduced computational overhead. 
Finally, a HiFi-GAN vocoder reconstructed time-domain waveforms from the refined mel-spectrogram to complete decoding. 
Extensive experiments on the LibriTTS dataset at 16~kHz and the VCTK dataset at 48~kHz demonstrated that, under the same ultra-low-bitrate constraints, FMelCodec consistently achieved stronger perceptual quality and speaker preservation than representative neural codec baselines, while maintaining a favorable bitrate–quality–complexity trade-off.
Future work will focus on further reducing vocoder-side computation, enabling low-latency streaming operation, and extending the CRR framework to broader speech coding scenarios.

\bibliographystyle{IEEEtran}
\bibliography{mybib}

@inproceedings{yang2024generative,
  title={Generative de-quantization for neural speech codec via latent diffusion},
  author={Yang, Haici and Jang, Inseon and Kim, Minje},
  booktitle={Proc. ICASSP},
  pages={1251--1255},
  year={2024},
  organization={IEEE}
}

@article{liu2024semanticodec,
  title={Semanticodec: An ultra low bitrate semantic audio codec for general sound},
  author={Liu, Haohe and Xu, Xuenan and Yuan, Yi and Wu, Mengyue and Wang, Wenwu and Plumbley, Mark D},
  journal={IEEE Journal of Selected Topics in Signal Processing},
  volume={18},
  number={8},
  pages={1448--1461},
  year={2024},
  publisher={IEEE}
}

@article{san2023discrete,
  title={From discrete tokens to high-fidelity audio using multi-band diffusion},
  author={San Roman, Robin and Adi, Yossi and Deleforge, Antoine and Serizel, Romain and Synnaeve, Gabriel and D{\'e}fossez, Alexandre},
  journal={Advances in neural information processing systems},
  volume={36},
  pages={1526--1538},
  year={2023}
}

@inproceedings{radford2023robust,
  title={Robust speech recognition via large-scale weak supervision},
  author={Radford, Alec and Kim, Jong Wook and Xu, Tao and Brockman, Greg and McLeavey, Christine and Sutskever, Ilya},
  booktitle={Proc. ICML},
  pages={28492--28518},
  year={2023},
}

@article{ho2020denoising,
  title={Denoising diffusion probabilistic models},
  author={Ho, Jonathan and Jain, Ajay and Abbeel, Pieter},
  journal={Advances in neural information processing systems},
  volume={33},
  pages={6840--6851},
  year={2020}
}

@inproceedings{pia2025flowmac,
  title={{FlowMAC}: Conditional flow matching for audio coding at low bit rates},
  author={Pia, Nicola and Strauss, Martin and Multrus, Markus and Edler, Bernd},
  booktitle={Proc. ICASSP},
  pages={1--5},
  year={2025},
}

@inproceedings{wu2025ts3,
  title={{TS3-Codec}: Transformer-Based Simple Streaming Single Codec},
  author={Wu, Haibin and Kanda, Naoyuki and Emre Eskimez, Sefik and Li, Jinyu},
  booktitle={Proc. Interspeech},
  pages={604--608},
  year={2025}
}

@inproceedings{parkerscaling,
  title={Scaling Transformers for Low-Bitrate High-Quality Speech Coding},
  author={Parker, Julian D and Smirnov, Anton and Pons, Jordi and Carr, CJ and Zukowski, Zack and Evans, Zach and Liu, Xubo},
  booktitle={The Thirteenth International Conference on Learning Representations}
}

@article{defossez2024moshi,
  title={Moshi: a speech-text foundation model for real-time dialogue},
  author={D{\'e}fossez, Alexandre and Mazar{\'e}, Laurent and Orsini, Manu and Royer, Am{\'e}lie and P{\'e}rez, Patrick and J{\'e}gou, Herv{\'e} and Grave, Edouard and Zeghidour, Neil},
  journal={arXiv preprint arXiv:2410.00037},
  year={2024}
}

@inproceedings{dellafocalcodec,
  title={{FocalCodec}: Low-Bitrate Speech Coding via Focal Modulation Networks},
  author={Della Libera, Luca and Paissan, Francesco and Subakan, Cem and Ravanelli, Mirco},
  booktitle={The Thirty-ninth Annual Conference on Neural Information Processing Systems}
}

@article{xu2024mucodec,
  title={Mucodec: Ultra low-bitrate music codec},
  author={Xu, Yaoxun and Chen, Hangting and Yu, Jianwei and Tan, Wei and Gu, Rongzhi and Lei, Shun and Lin, Zhiwei and Wu, Zhiyong},
  journal={arXiv preprint arXiv:2409.13216},
  year={2024}
}

@inproceedings{jiang2024mdctcodec,
  title={{MDCTC}odec: A Lightweight {MDCT}-Based Neural Audio Codec Towards High Sampling Rate and Low Bitrate Scenarios},
  author={Jiang, Xiao-Hang and Ai, Yang and Zheng, Rui-Chen and Du, Hui-Peng and Lu, Ye-Xin and Ling, Zhen-Hua},
  booktitle={Proc. SLT},
  pages={540--547},
  year={2024}
}

@article{yamagishi2019cstr,
  title={{CSTR VCTK} Corpus: English multi-speaker corpus for {CSTR} voice cloning toolkit (version 0.92)},
  author={Yamagishi, Junichi and Veaux, Christophe and MacDonald, Kirsten and others},
  journal={University of Edinburgh. The Centre for Speech Technology Research (CSTR)},
  pages={271--350},
  year={2019}
}

@inproceedings{loshchilovdecoupled,
  title={Decoupled Weight Decay Regularization},
  author={Loshchilov, Ilya and Hutter, Frank},
  booktitle={Proc. ICLR},
year={2017}
}

@inproceedings{kleijn2021generative,
  title={Generative speech coding with predictive variance regularization},
  author={Kleijn, W Bastiaan and Storus, Andrew and Chinen, Michael and Denton, Tom and Lim, Felicia SC and Luebs, Alejandro and Skoglund, Jan and Yeh, Hengchin},
  booktitle={Proc. ICASSP},
  pages={6478--6482},
  year={2021},
}

@article{zheng2025ervq,
  title={E{RVQ}: Enhanced residual vector quantization with intra-and-inter-codebook optimization for neural audio codecs},
  author={Zheng, Rui-Chen and Du, Hui-Peng and Jiang, Xiao-Hang and Ai, Yang and Ling, Zhen-Hua},
  journal={IEEE Transactions on Audio, Speech and Language Processing},
  year={2025},
  volume={33},
  pages={2539--2550},
}

@inproceedings{chinen2020visqol,
  title={Vi{SQOL} v3: An open source production ready objective speech and audio metric},
  author={Chinen, Michael and Lim, Felicia SC and Skoglund, Jan and Gureev, Nikita and O’Gorman, Feargus and Hines, Andrew},
  booktitle={Proc. QoMEX},
  pages={1--6},
  year={2020}
}

@article{ai2024apcodec,
  title={{APCodec: A} Neural Audio Codec with Parallel Amplitude and Phase Spectrum Encoding and Decoding},
  author={Ai, Yang and Jiang, Xiao-Hang and Lu, Ye-Xin and Du, Hui-Peng and Ling, Zhen-Hua},
  journal={IEEE/ACM Transactions on Audio, Speech, and Language Processing},
  volume={32},
  pages={3256--3269},
  year={2024}
}

@ARTICLE{zeghidour2021soundstream,
  author={Zeghidour, Neil and Luebs, Alejandro and Omran, Ahmed and Skoglund, Jan and Tagliasacchi, Marco},
  journal={IEEE/ACM Transactions on Audio, Speech, and Language Processing}, 
  title={{SoundStream: An End-to-End Neural Audio Codec}}, 
  year={2022},
  volume={30},
  number={},
  pages={495-507},
}

@ARTICLE{11303581,
  author={Liu, Fei and Ai, Yang and Lu, Ye-Xin and Zheng, Rui-Chen and Du, Hui-Peng and Ling, Zhen-Hua},
  journal={IEEE Transactions on Audio, Speech and Language Processing}, 
  title={Universal Discrete-Domain Speech Enhancement}, 
  year={2026},
  volume={34},
  number={},
  pages={285-298},}

@article{defossez2023high,
  title={High Fidelity Neural Audio Compression},
  author={D{\'e}fossez, Alexandre and Copet, Jade and Synnaeve, Gabriel and Adi, Yossi},
  journal={Transactions on Machine Learning Research},
  year={2023}
}

@inproceedings{lv2024freev,
  title     = {{FreeV: Free Lunch For Vocoders Through Pseudo Inversed Mel Filter}},
  author={Lv, Yuanjun and Li, Hai and Yan, Ying and Liu, Junhui and Xie, Danming and Xie, Lei},
  year      = {2024},
  booktitle = {Proc. {Interspeech}},
  pages     = {3869--3873},
}

@article{langman2024spectral,
  title={Spectral codecs: Spectrogram-based audio codecs for high quality speech synthesis},
  author={Langman, Ryan and Juki{\'c}, Ante and Dhawan, Kunal and Koluguri, Nithin Rao and Ginsburg, Boris},
  journal={arXiv preprint arXiv:2406.05298},
  year={2024}
}

@inproceedings{lipman2023flow,
  title={Flow Matching for Generative Modeling},
  author={Lipman, Yaron and Chen, Ricky TQ and Ben-Hamu, Heli and Nickel, Maximilian and Le, Matt},
  booktitle={Proc. ICLR},
  year={2023}
}

@article{jiang2025streamable,
  title={A Streamable Neural Audio Codec With Residual Scalar-Vector Quantization for Real-Time Communication},
  author={Jiang, Xiao-Hang and Ai, Yang and Zheng, Rui-Chen and Ling, Zhen-Hua},
  journal={IEEE Signal Processing Letters},
  volume={32},
  pages={1645--1649},
  year={2025}
}

@article{vaswani2017attention,
  title={Attention is all you need},
  author={Vaswani, Ashish and Shazeer, Noam and Parmar, Niki and Uszkoreit, Jakob and Jones, Llion and Gomez, Aidan N and Kaiser, {\L}ukasz and Polosukhin, Illia},
  journal={Advances in neural information processing systems},
  volume={30},
  year={2017}
}

@article{he2016identity,
  title={Identity Mappings in Deep Residual Networks},
  author={He, Kaiming and Zhang, Xiangyu and Ren, Shaoqing and Sun, Jian},
  journal={Computer Vision--ECCV 2016},
  volume={9908},
  pages={630--645},
  year={2016},
  publisher={Springer International Publishing}
}

@inproceedings{li2024single,
  title={Single-Codec: Single-Codebook Speech Codec towards High-Performance Speech Generation},
  author={Li, Hanzhao and Xue, Liumeng and Guo, Haohan and Zhu, Xinfa and Lv, Yuanjun and Xie, Lei and Chen, Yunlin and Yin, Hao and Li, Zhifei},
  booktitle={Proc. Interspeech},
  pages={3390--3394},
  year={2024}
}

@article{chen2022wavlm,
  title={{WavLM}: Large-scale self-supervised pre-training for full stack speech processing},
  author={Chen, Sanyuan and Wang, Chengyi and Chen, Zhengyang and Wu, Yu and Liu, Shujie and Chen, Zhuo and Li, Jinyu and Kanda, Naoyuki and Yoshioka, Takuya and Xiao, Xiong and others},
  journal={IEEE Journal of Selected Topics in Signal Processing},
  volume={16},
  number={6},
  pages={1505--1518},
  year={2022},
  publisher={IEEE}
}

@article{van2017neural,
  title={Neural discrete representation learning},
  author={Van Den Oord, Aaron and Vinyals, Oriol and others},
  journal={Advances in neural information processing systems},
  volume={30},
  year={2017}
}

@inproceedings{zheng2023online,
  title={Online clustered codebook},
  author={Zheng, Chuanxia and Vedaldi, Andrea},
  booktitle={Proc. ICCV},
  pages={22798--22807},
  year={2023}
}

@article{baevski2020wav2vec,
  title={{wav2vec 2.0}: A framework for self-supervised learning of speech representations},
  author={Baevski, Alexei and Zhou, Yuhao and Mohamed, Abdelrahman and Auli, Michael},
  journal={Advances in neural information processing systems},
  volume={33},
  pages={12449--12460},
  year={2020}
}

@inproceedings{popov2021grad,
  title={{Grad-TTS}: A diffusion probabilistic model for text-to-speech},
  author={Popov, Vadim and Vovk, Ivan and Gogoryan, Vladimir and Sadekova, Tasnima and Kudinov, Mikhail},
  booktitle={Proc. ICML},
  pages={8599--8608},
  year={2021},
}

@article{razavi2019generating,
  title={Generating diverse high-fidelity images with vq-vae-2},
  author={Razavi, Ali and Van den Oord, Aaron and Vinyals, Oriol},
  journal={Advances in neural information processing systems},
  volume={32},
  year={2019}
}

@inproceedings{wang2025flowse,
  title     = {Flow{SE}: Efficient and High-Quality Speech Enhancement via Flow Matching},
  author={Wang, Ziqian and Liu, Zikai and Zhu, Xinfa and Zhu, Yike and Liu, Mingshuai and Chen, Jun and Xiao, Longshuai and Weng, Chao and Xie, Lei},
  year      = {2025},
  booktitle = {Proc. {Interspeech}},
  pages     = {4858--4862},
}

@inproceedings{welkerflowdec,
  title={{FlowDec}: A flow-based full-band general audio codec with high perceptual quality},
  author={Welker, Simon and Le, Matthew and Chen, Ricky TQ and Hsu, Wei-Ning and Gerkmann, Timo and Richard, Alexander and WU, YI-CHIAO},
  booktitle={Proc. ICLR},
year={2025}
}

@inproceedings{kumar2024high,
  title={High-fidelity audio compression with improved rvqgan},
  author={Kumar, Rithesh and Seetharaman, Prem and Luebs, Alejandro and Kumar, Ishaan and Kumar, Kundan},
  booktitle={Proc. NeurIPS},
  volume={36},
  year={2024}
}

@inproceedings{jiwavtokenizer,
  title={{WavTokenizer}: an Efficient Acoustic Discrete Codec Tokenizer for Audio Language Modeling},
  author={Ji, Shengpeng and Jiang, Ziyue and Wang, Wen and Chen, Yifu and Fang, Minghui and Zuo, Jialong and Yang, Qian and Cheng, Xize and Wang, Zehan and Li, Ruiqi and others},
  booktitle={Proc. ICLR},
year={2025}
}

@article{xin2024bigcodec,
  title={{BigCodec}: Pushing the limits of low-bitrate neural speech codec},
  author={Xin, Detai and Tan, Xu and Takamichi, Shinnosuke and Saruwatari, Hiroshi},
  journal={arXiv preprint arXiv:2409.05377},
  year={2024}
}

@article{chen2024vall,
  title={{VALL-E}2: Neural codec language models are human parity zero-shot text to speech synthesizers},
  author={Chen, Sanyuan and Liu, Shujie and Zhou, Long and Liu, Yanqing and Tan, Xu and Li, Jinyu and Zhao, Sheng and Qian, Yao and Wei, Furu},
  journal={arXiv preprint arXiv:2406.05370},
  year={2024}
}

@inproceedings{dietz2015overview,
  title={Overview of the {EVS} codec architecture},
  author={Dietz, Martin and Multrus, Markus and Eksler, Vaclav and Malenovsky, Vladimir and Norvell, Erik and Pobloth, Harald and Miao, Lei and Wang, Zhe and Laaksonen, Lasse and Vasilache, Adriana and others},
  booktitle={Proc. ICASSP},
  pages={5698--5702},
  year={2015},
}

@inproceedings{valin2013high,
  title={High-Quality, Low-Delay Music Coding in the Opus Codec},
  author={Valin, Jean-Marc and Maxwell, Gregory and Terriberry, Timothy B and Vos, Koen},
  booktitle={Audio Engineering Society Convention 135},
  year={2013},
  organization={Audio Engineering Society}
}

@inproceedings{goodfellow2014generative,
	title={Generative adversarial nets},
	author={Goodfellow, Ian and Pouget-Abadie, Jean and Mirza, Mehdi and Xu, Bing and Warde-Farley, David and Ozair, Sherjil and Courville, Aaron and Bengio, Yoshua},
	booktitle={Proc. NeurIPS},
	volume={27},
	year={2014}
}

@inproceedings{du2023apnet2,
  title={{APNet2: High-quality and high-efficiency neural vocoder with direct prediction of amplitude and phase spectra}},
  author={Du, Hui-Peng and Lu, Ye-Xin and Ai, Yang and Ling, Zhen-Hua},
  booktitle={Proc. NCMMSC},
  pages={66--80},
  year={2023},
}

@inproceedings{woo2023convnext,
	title={{ConvNeXt v2: Co-designing and scaling convnets with masked autoencoders}},
	author={Woo, Sanghyun and Debnath, Shoubhik and Hu, Ronghang and Chen, Xinlei and Liu, Zhuang and Kweon, In So and Xie, Saining},
	booktitle={Proc. CVPR},
	pages={16133--16142},
	year={2023}
}

@inproceedings{saeki2022utmos,
  title={{UTMOS: UTokyo-SaruLab System for VoiceMOS Challenge 2022}},
  author={Saeki, Takaaki and Xin, Detai and Nakata, Wataru and Koriyama, Tomoki and Takamichi, Shinnosuke and Saruwatari, Hiroshi},
  booktitle={Proc. Interspeech},
  pages={4521--4525},
  year={2022},
}

@inproceedings{siuzdak2023vocos,
  title={Vocos: Closing the gap between time-domain and Fourier-based neural vocoders for high-quality audio synthesis},
  author={Siuzdak, Hubert},
  booktitle={Proc. ICLR},
    year={2024}
}

@article{zen2019libritts,
  title={Libritts: A corpus derived from librispeech for text-to-speech},
  author={Zen, Heiga and Dang, Viet and Clark, Rob and Zhang, Yu and Weiss, Ron J and Jia, Ye and Chen, Zhifeng and Wu, Yonghui},
  journal={arXiv preprint arXiv:1904.02882},
  year={2019}
}

@inproceedings{lee2023bigvgan,
  title={{BigVGAN}: A UNIVERSAL NEURAL VOCODER WITH LARGE-SCALE TRAINING},
  author={Lee, Sang Gil and Ping, Wei and Ginsburg, Boris and Catanzaro, Bryan and Yoon, Sungroh},
  booktitle={Proc. ICLR},
  year={2023}
}

@inproceedings{li2025speech,
  title={Speech enhancement using continuous embeddings of neural audio codec},
  author={Li, Haoyang and Yip, Jia Qi and Fan, Tianyu and Chng, Eng Siong},
  booktitle={Proc. ICASSP},
  pages={1--5},
  year={2025},
}

@ARTICLE{10842513,
  author={Chen, Sanyuan and Wang, Chengyi and Wu, Yu and Zhang, Ziqiang and Zhou, Long and Liu, Shujie and Chen, Zhuo and Liu, Yanqing and Wang, Huaming and Li, Jinyu and He, Lei and Zhao, Sheng and Wei, Furu},
  journal={IEEE Transactions on Audio, Speech and Language Processing}, 
  title={Neural Codec Language Models are Zero-Shot Text to Speech Synthesizers}, 
  year={2025},
  volume={33},
  number={},
  pages={705-718},
}

@inproceedings{mehta2024matcha,
  title={{Matcha-TTS}: A fast TTS architecture with conditional flow matching},
  author={Mehta, Shivam and Tu, Ruibo and Beskow, Jonas and Sz{\'e}kely, {\'E}va and Henter, Gustav Eje},
  booktitle={Proc. ICASSP},
  pages={11341--11345},
  year={2024},
}

@inproceedings{kong2020hifi,
	title={{HiFi-GAN}: Generative adversarial networks for efficient and high fidelity speech synthesis},
	author={Kong, Jungil and Kim, Jaehyeon and Bae, Jaekyoung},
	booktitle={Proc. NeurIPS},
	volume={33},
	pages={17022--17033},
	year={2020}
}
\end{document}